\documentclass[amsmath,amssymb,epsfig,aps,pra,twocolumn]{revtex4-1}
\usepackage{amsmath}
\usepackage{epsfig}
\usepackage{graphicx}
\usepackage{color}
\usepackage{amssymb}
\usepackage{epstopdf}
\usepackage[utf8]{inputenc}
\usepackage[T1]{fontenc}
\begin{document}
\title{Spatial separation of rotating binary Bose-Einstein condensate by tuning the dipolar 
interactions}\thanks{Corresponding author: tomio@ift.unesp.br}
\author{Ramavarmaraja Kishor Kumar$^1$, Lauro Tomio$^{2,3}$, and Arnaldo Gammal$^1$}
\affiliation{$^{1}$Instituto de F\'{i}sica, Universidade de S\~{a}o Paulo, 05508-090 S\~{a}o Paulo, Brazil.\\
$^{2}$Instituto de F\'isica Te\'orica, Universidade Estadual Paulista, 01156-970 S\~ao Paulo, SP, Brazil.\\
$^{3}$Instituto Tecnol\'ogico de Aeron\'autica, DCTA,12.228-900 S\~ao Jos\'e dos Campos, SP, Brazil.}
\date{\today}
\begin{abstract}
We are pointing out relevant anisotropic 
effects, related to spatial separation, miscibility and mass-symmetry, 
due to dipole-dipole interactions in rotating binary dipolar Bose-Einstein condensates,
by considering symmetric  ($^{164}$Dy-$^{162}$Dy) and asymmetric  ($^{168}$Er-$^{164}$Dy, 
$^{164}$Dy-$^{87}$Rb) dipolar mixtures. The binary mixtures are kept in strong pancake-shaped trap, modeled 
by an effective two-dimensional coupled Gross-Pitaevskii equation.  
The anisotropy of the dipolar interactions, on miscibility and vortex-lattice structures, is studied by   
tuning the polarization angle of the dipoles $\varphi$,  which can enhance the attractive part of the 
dipole-dipole interaction (DDI) for both inter- and intra-species. Within this procedure of changing to attractive 
the DDI, a clear spatial  separation is verified in the densities at some critical polarization angle.
The spatial separations, being angular for symmetric mixtures and radial for asymmetric ones, are
verified for repulsive contact interactions when the inter- to intra-species ratio $\delta$ is larger than one,
implying the system is less miscible. The corresponding result for the critical polarization angle as a function 
of $\delta$ is shown in the particular dipolar symmetric case. A striking outcome of the present study is the 
observed sensibility of the vortex-pattern binary distributions due to the mass-asymmetry. This is exemplified by the 
symmetric dipolar mixture, where the two isotopes are of the same species. 
\end{abstract}

\flushbottom

\maketitle
\section{Introduction}
The realization of Bose-Einstein condensation with chromium ($^{52}$Cr) atoms has opened the new research 
direction called dipolar quantum gases~\cite{2005-Pfau},  allowing first experimental 
studies on strong dipolar effects in quantum superfluid~\cite{2007-ferrofluid}. 
Following these investigations with chromium, many subsequent studies have been carried out by different 
experimental groups on fermionic and bosonic properties of strongly dipolar ultracold gases, such as with dysprosium and 
erbium~\cite{Lu2010,Aikawa2012,2017-Ulitzsch,2018-natphys-Ferlaino,2018PRA97-Ferlaino2,2018-Grimm,2018NJP-Velji,2018arxiv-Ferlaino}.
Also recently, it was reported in Ref.~\cite{2018PRA97-Ferlaino} the realization of two-species mixtures with
strongly dipolar atoms, using magneto-optical trap with erbium and dysprosium. 
By improving the experimental production of dysprosium ($^{162}$Dy) Bose-Einstein condensates, a new technique 
was reported in Ref.~\cite{Lucioni2018}, which allows efficient loading from a magneto-optical trap. 
Following that, the ability to tune the strength of the dipole-dipole interaction (DDI) has been investigated in Refs.~\cite{Tang2018,Cai2018}, for
rotating strongly-dipolar single-atom species. These highly magnetic lanthanide atoms, having strong DDIs, can present quite relevant 
and interesting quantum behaviors, including ferro-fluidity and self-bound 
droplets~\cite{2007-ferrofluid,2016-droplet,2016-Ferlaino}. 
As demonstrated in Refs.~\cite{2016-droplet}, the quantum fluctuations in strongly dipolar Bose gases can stabilize 
droplets against the mean-field collapse. 
A recent review considering self-bound droplets can be found in Ref.~\cite{2018-yukalov}.
Related to vortices in rotating dipolar Bose-Einstein condensates, contrasting with non-dipolar systems, 
where it is being explored the interplay of magnetism with vorticity, a review is given in Ref.~\cite{2017-Martin}.
Novel fascinating possibilities in physics are being revealed by the researchers with strongly dipolar binary 
mixtures~\cite{2018PRA97-Ferlaino}, due to the peculiar competition between isotropic short-range contact interaction 
and long-range anisotropic DDI. 
Apart from experimental studies, there are previous investigations on the stability of trapped dipolar Bose-Einstein condensates
in Ref.~\cite{2003-Santos}, as well as several theoretical studies performed with dipolar mean-field theory, which are based on 
the construction of the corresponding pseudo-potential~\cite{2004-MFdipolar,2001-Yi}. 
Following that, on binary dipolar mixtures, one can also find out more recent theoretical investigations as, for example, in
Ref.~\cite{2011-Zaman}, where it was studied the effect of a second dipolar BEC on the stability properties of binary dipolar mixtures 
in pancake-trapped symmetry.
The ferrofluid-like pattern formations are studied in two-component BEC with DDI 
in Ref.~\cite{Saito2009}, with instabilities and pattern formations (verified by oppositely polarized dipoles in 
a two-component BEC) being recently reported in Ref.~\cite{Saito2018}.
Rotational properties of two-component dipolar BEC in concentrically coupled annular traps were also
studied in Ref.~\cite{2015Zhang}, by assuming one component without dipole moments.
The immiscibility-miscibility transition of a two-component dipolar BECs, previously studied in Ref.~\cite{Wilson2012},
was further investigated by some of us in Ref.~\cite{2017jpco}, where it was determined the miscibility properties
of binary mixtures with the isotopes $^{162,164}$Dy and $^{168}$Er within a three-dimensional (3D) 
formalism.  When considering the miscibility, it was also shown in Ref.~\cite{Kumar2017} that the rotational properties of 
symmetric and asymmetric dipolar mixtures present quite distinct vortex lattice structures. 

In the present work we investigate properties of strong dipolar mixtures in a two-dimensional (2D) rotating magnetic trap,
by tuning the dipole-dipole interaction, adjusting the polarization angle between dipoles, within a
coupled Gross-Pitaevskii (GP) formalism.  In view of previous stability analysis and analytical considerations 
on miscibility and expected vortex-pattern structures, our approach is fully performed by exact numerical solutions of the
corresponding formalism.
These studies are motivated by the recent improvements in the experimental 
techniques~\cite{2018PRA97-Ferlaino}, in order to control the dipole orientations of dipolar condensed atoms and explore 
new physics. By tuning the dipoles larger than the angle aligned along the $z$-axis provides an attractive component in the DDI. 
Since, the self-bound droplets require a balance between repulsive and attractive interactions, by controlling 
the dipole orientations it is expected to reach more favorable conditions for the observation of self-bound droplets. 
The stability of dipolar BECs affected by the attractive part of the DDI can be kept by 
using strong axially confining pancake-shaped trap, with suitable repulsive contact interactions. 
Therefore, within a study which can be easily extended to other binary mixtures, we consider in particular 
atomic species having strong magnetic dipolar properties, such as the ones used in the 
recent BEC  experiments reported in Refs.~\cite{2018PRA97-Ferlaino,2018PRA97-Ferlaino2}.
So, our main focus are the binary mixtures having the isotopes of dysprosium $^{162,164}$Dy and erbium $^{168}$Er.
For comparison, in our analysis we also include a binary mixture in which one of the species is the weakly dipolar 
rubidium $^{87}$Rb.

We assume a configuration in which the dipoles are oriented along the $z-$axis, within a quasi-2D settings where 
the dipoles are parallel to each other in the $(x,y)$-plane.  By tuning the dipoles, exploring anisotropic effects due to 
DDI in the $(x,y)$-plane, different properties are evidenced for the rotating binary mixtures. 
Using this procedure to alter the DDI from repulsive to attractive, one can further explore condensate 
properties of coupled systems for different contact interactions. 
Some of the relevant properties of the strong dipolar binary systems considered in this work ($^{164}$Dy-$^{162}$Dy, 
$^{168}$Er-$^{164}$Dy, and $^{164}$Dy-$^{87}$Rb), have been studied recently.
In Ref.~\cite{2017jpco}, within a full 3D formalism, it was established the miscible-immiscible stable conditions of these 
coupled dipolar systems with repulsive contact interactions, from pancake- to cigar-type trap configurations.
The rotational properties and vortex-lattice structures were further investigated in 
Refs.~\cite{Kumar2017,2016Xiao}. More recently, in Ref.~\cite{Kumar2018}, we have also investigated the effect 
of changing the inter- to intra-species scattering length in these strong dipolar systems under squared optical lattices.
These mentioned works helped us to better define some characteristics of such dipolar binary systems which can 
be relevant for experimental realization. Among these characteristics, one should noticed the possibility of tuning the 
polarization angles of both interacting dipoles, such that the effective time-averaged dipole-dipole potential can be 
modified from repulsive to attractive interactions. Therefore, it is expected that the interplay between DDIs and 
contact interactions can bring us some possible interesting effect in the rotational properties of binary dipolar systems. 
Within such motivation, by considering the above strongly dipolar binary systems kept in quasi-2D pancake-shaped traps,
our main task is to investigate possible anisotropic effects due to dipole-dipole interactions in rotating 
binary dipolar Bose-Einstein condensates. In this case, by assuming fixed the rotation frequency and trap aspect ratio,
we have two independent mechanisms to control de inter- and intra-species interactions: Feshbach resonance  
techniques~\cite{1998inouye}, controlling the contact interactions; and the tuning of the polarization angle, modifying 
the dipole-dipole interactions.

By considering the interplay between the contact and dipolar interactions, 
some peculiar non-trivial behaviors emerge in the density distributions, which for symmetric 
mixtures such as  $^{164}$Dy-$^{162}$Dy are quite distinct than the corresponding distributions for non-symmetric 
ones as $^{168}$Er-$^{164}$Dy, and $^{164}$Dy-$^{87}$Rb. 
Symmetric dipolar mixtures with isotopes of the same atomic species are quite interesting to study due to the rich 
variety of patterns that one can create in the density distributions, from geometric vortex structures to droplet-like 
formations. Interesting enough, it also leads to an almost complete half-space angular separation between
the species, for some critical polarization angle when the DDI is attractive.
The observation of a critical polarization angle, at which an almost complete spatial separation occurs between the 
two densities when the DDI is attractive,  is quite remarkable for both symmetric and asymmetric mixtures. 
Being half-space angular for the dipolar symmetric mixture, the spatial separation of the densities turns out to be 
radial in case of asymmetric mixtures. 
Beyond that, the particular case of dipolar-symmetric mixture provides us the opportunity to verify the effect of 
mass-symmetry breaking in the two-component vortex-pattern structures. When considering a small change in 
the mass, the sensitivity of the patterns is evidenced with a fixed rotation parameter and by adjusting to the 
same value the inter- and intra-species contact interactions.

Next, we present the basic formalism and notation, followed by a section with our main results
for the three kind of dipolar mixtures we consider. In a final section, we have a summary with our main conclusions.

\section{Formalism and parametrization}
\label{secII} 
\begin{figure}[tbp]
\centerline{
\includegraphics[width=0.5\textwidth]{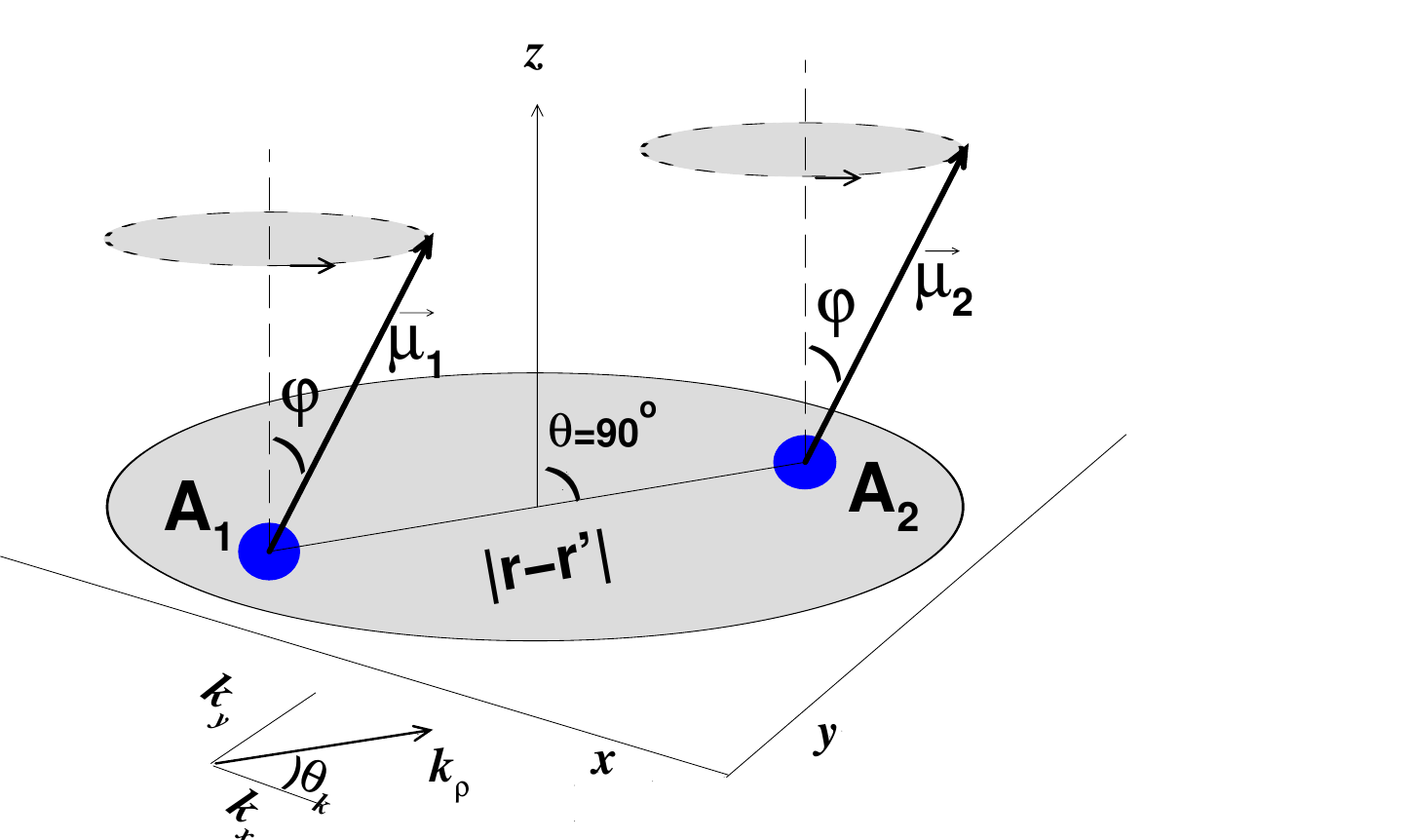}
}
\caption{
An illustration of the DDI between the two atomic species 1 and 2, given in coordinate space, 
where the polarization angle is $\varphi$, with the assumption that both atoms are in the $(x,y)$ plane 
($\theta=90^o$). 
}
\label{fig01}
\end{figure}

The coupled dipolar system with condensed two atomic species are assumed to be confined in strongly pancake-shaped 
harmonic traps, with fixed aspect ratios, such that $\lambda=\omega_{i,z}/\omega_{i,\perp}=20$ for both species $i=1,2$, 
where $\omega_{i,z}$ and $\omega_{i,\perp}$ are, respectively, the longitudinal and transverse trap frequencies.
The coupled Gross-Pitaevskii (GP) equation is  cast in a dimensionless format by measuring the energy  in units of 
$\hbar \omega_{1,\perp}$, with the length in units of $l_\perp\equiv \sqrt{\hbar/(m_1\omega_{1,\perp})}$, where 
$\omega_{1,\perp}\equiv \omega_{1}$ and $m_1$ are the transverse frequency and mass for the species $i=1$. 
Correspondingly, the space and time variables are given in units of $l_\perp$ and $1/\omega_{1}$, respectively, 
such that  ${\bf r}\to l_\perp {\bf r}$ and $t\to \tau/\omega_{1}$).  Within these units, and 
by adjusting both trap frequencies such that ${m_2\omega_{2,\perp}^2}={m_1\omega_{1,\perp}^2}$, 
the dimensionless external three-dimensional (3D) trap potential for both species can be written as 
{\small\begin{equation}
V_{3D}({\bf r})=\frac{1}{2}(x^{2}+y^{2}+\lambda^2 z^2)\equiv V(x,y) +\frac{1}{2}\lambda^2z^2.
\label{3Dtrap}
\end{equation}}  
In the case of the DDI, we consider that one of the atoms (A$_i$) is at position ${\bf r}$, with the other one (A$_j$)
at position ${\bf r^\prime}$, respectively, with their line-vector ${\bf r-r^\prime}$ making an angle $\theta$ related to the $z-$axis, such 
that $\theta\approx 90^\circ$ for a strong pancake-shaped trap as given by Eq.~(\ref{3Dtrap}) 
(each one of the atoms can be of species $i,j=1,2$).
Next, we assume both dipoles are polarized in the same direction, making an angle $\varphi$ with respect to the $z-$axis.
See the corresponding illustration in the Fig.~\ref{fig01}. 
The tunability is performed by using time-dependent magnetic fields with dipoles rapidly rotating around the 
$z-$axis~\cite{goral2002,2002Pfau} . 
The magnetic field is given by the combination of a static part along the $z-$direction and a fast rotating part in the 
$(x,y)-$plane, having a frequency such that the atoms are not significantly moving during each period. 
Once performed a time averaging of the DDI within a period, the corresponding 3D averaged interaction  
between the coupled dipolar species $i$ and $j$, with their respective magnetic dipole moments $\mu_i$ 
and $\mu_j$ given in terms of the Bohr magneton $\mu_B$, can be written in our dimensionless format as~\cite{2002Pfau} 
{\small \begin{equation}
\left<V^{(d)}_{3D}({{\bf r}-{\bf r}^\prime})\right>_{ij}=\frac{\mu_{0}\mu_{i}\mu _{j}}{\hbar\omega_1 l_\perp^3}
\frac{1-3\cos ^{2}\theta}{\left|{{\bf r}-{\bf r}^\prime}\right|^{3}}
\left(\frac{{3\cos ^{2}\varphi}-1}{2}\right), 
\label{DDI}
\end{equation}} 
where $\mu _{0}$ is the free-space permeability.
In the following, we assume $\theta=90^\circ$, such that the DDI strength is repulsive for $\varphi<\varphi_M$ and 
attractive when $\varphi>\varphi_M$ (where $\varphi_M\approx 54.7^\circ$ is the so-called ``magic angle", when the DDI 
is averaged to zero). The factor within parenthesis in Eq.~(\ref{DDI}) results from the time-averaging procedure on 
the dipole orientation around the $z-$axis. 
A large value for the aspect ratio $\lambda$ allows us to reduce the original 3D formalism to a 2D, by considering
the usual factorization of the 3D wave function into the 2D wave function $\psi_{i}(x,y,\tau)$ and the ground state of 
the transverse harmonic oscillator trap, given by the ansatz
$\left({\lambda}/{\pi}\right)^{1/4}e^{-{\lambda z^{2}}/{2}}$.
The two-body contact interactions related to the scattering lengths $a_{ij}$, as well as the dipole-dipole interaction 
parameters for the species $i,j = 1,2$, are defined as~\cite{2017jpco}
\begin{eqnarray}
g_{ij}&\equiv& \sqrt{2\pi \lambda }\frac{(m_{1}+m_2)}{m_2}\frac{a_{ij}N_{j}}{l_\perp},\;\;
\mathrm{d}_{ij}=\frac{N_{i}}{4\pi }\frac{\mu_{0}\mu_{i}\mu _{j}}{\hbar \omega_{1}\,l_\perp^{3}},
\;\; \nonumber\\
a_{ii}^{(d)}&\equiv& \frac{1}{12\pi }\frac{m_{i}}{m_{1}}\frac{\mu_{0}\mu _{i}^2}{\hbar \omega_{1}l_\perp^{2}},
\;\; a_{12}^{(d)}=a_{21}^{(d)}=\frac{1}{12\pi }\frac{\mu_{0}\mu_{1}\mu _{2}}{\hbar \omega_{1}l_\perp^{2}},
\label{par}
\end{eqnarray}
where $N_{j=1,2}$ is the number of atoms of the species $j$. 
In the next, the length unit will be adjusted to  $l_\perp = 1\mu$m$ \approx 1.89\times 10^4 a_0$, where 
$a_0$ is the Bohr radius, such that $a_{ij}$ and $a_{ij}^{(d)}$ can be conveniently given in terms of $a_0$ in our numerical analysis.  
   
With the above notations for units and parameters, the corresponding coupled GP equation in 2D is given by 
{\small \begin{eqnarray}
\mathrm{i}\frac{\partial \psi_{i}  }{\partial \tau }
&=&{\bigg [}\frac{-m_{1}}{2m_{i}}{\left(\frac{\partial^2}{\partial x^2}+\frac{\partial^2}{\partial y^2}\right)}
+V(x,y)- \Omega L_{z}+\sum_{j=1,2}g_{ij}|\psi_{j} |^{2}
\nonumber\\
&+&\sum_{j=1,2}  \mathrm{d}_{ij} \int_{-\infty}^{\infty}d x^\prime d y^\prime V^{(d)}(x-x^\prime,y-y^\prime)
|\psi_{j}^\prime  |^{2}
 {\bigg ]}
\psi_{i}  
\label{2d-2c}, 
\end{eqnarray}}
where $V^{(d)}(x,y)$ is the reduced 2D expression for the DDI, and 
$\psi_{i}\equiv \psi_{i}(x,y,\tau)$ and $\psi_{i}^\prime\equiv \psi_{i}(x^\prime,y^\prime,\tau)$
are the components of the total 2D wave function, normalized to one, 
$\int_{-\infty}^{\infty}dxdy|\psi _{i}|^{2}=1.$
$L_z$ is the angular momentum operator with $\Omega$ the corresponding rotation parameter (in units of  $\omega_{1}$), 
which is common for the two components. The external potential provided by the harmonic trap is given by 
$V(x,y)=\frac{1}{2}(x^{2}+y^{2})$.

In 2D momentum space, the DDI can be expressed as the combination of two terms, considering the orientations of the 
dipoles $\varphi$ and projection of the Fourier transformed of $V^{(d)}(x,y)$ in momentum space; one term  
perpendicular with the other parallel to the direction of the dipole inclinations, respectively given by~\cite{Wilson2012,2016Xiao}
{\small \begin{eqnarray}
\widetilde{V}_\perp^{(d)}( k_x, k_y) &=& 2 - 3 \sqrt{\frac{\pi}{2\lambda}}k_{\rho}
\exp \left( \frac{k_{\rho }^{2}}{2\lambda}\right) {\rm erfc}\left( \frac{k_{\rho }}{\sqrt{2\lambda}}\right)
,\label{DDI-perp}\\
\widetilde{V}_\parallel^{(d)}(k_x, k_y) &=& -1 + 3\frac{k_x^2}{k_\rho} \sqrt{\frac{\pi}{2\lambda}}
\exp \left( \frac{k_{\rho }^{2}}{2\lambda}\right) 
{\rm erfc}\left( \frac{k_{\rho }}{\sqrt{2\lambda}}\right)\label{DDI-par},
\end{eqnarray} }
where $ k_\rho^2\equiv k_x^2+k_y^2$, with $\rm erfc(x)$ being the complementary error function of $x$. 
The $k_x$ explicit in the right-hand-side of the above parallel term is the projected wave-number in the $(x,y)$ plane, in the 
direction of the polarization tilt, which was arbitrarily assumed in the $x-$axis when derived that term~\cite{Wilson2012}.
Generalizing the description to a polarization field rotating in the $(x,y)$ plane, as 
${\bf k_\rho}=(k_x,k_y)=(k_\rho\cos\theta_k, k_\rho\sin\theta_k)$ and all directions $\theta_k$
are equally possible, we should average $k_\rho^2\cos^2\theta_k$ in the plane, such that 
$k_x^2$ should be replaced by $k_\rho^2/2$ in the parallel term shown in Eq.~(\ref{DDI-par}). 
Therefore, by combining the two terms according to the orientations $\varphi$ of the
dipoles, we obtain the total 2D momentum-space DDI, 
$\widetilde{V}^{(d)}(k_x,k_y)=\cos^2(\varphi) \widetilde{V}_\perp^{(d)}(k_x,k_y) 
+ \sin^2(\varphi) \widetilde{V}_\parallel^{(d)}(k_x,k_y)$, which can be written as 
{\small \begin{eqnarray}
\widetilde{V}^{(d)}( k_x, k_y) &=& \frac{3\cos^2(\varphi)-1}{2}
\widetilde{V}_\perp^{(d)}( k_x, k_y) 
 \equiv V_\varphi(k_\rho)
\label{DDI-2D}.
\end{eqnarray} 
}
It follows that the effective DDI in 2D configuration space shown in Eq.~(\ref{2d-2c}) is obtained by 
applying the convolution theorem, performing the inverse 2D Fourier-transform for the product of the DDI 
and density, such that $\int dx^\prime dy^\prime V^{(d)}(x-x^{\prime },y-y^{\prime })|\psi _{j}^{\prime}|^{2}
=\mathcal{F}_{2D}^{-1}\left[ \widetilde{{V}}^{(d)}(k_x,k_y){\widetilde{n}_{j}}(k_x,k_y)\right]. $
From Eqs.~(\ref{DDI-perp}) and (\ref{DDI-2D}), one should noticed that such momentum-space Fourier transform of the 
dipole-dipole potential is changing the signal at some particular large momentum $k_\rho$. However,
after applying convolution theorem with the inverse Fourier transform (by integrating the momentum variables), 
the corresponding coordinate-space interaction has a definite value, as in the 3D case, which is positive for 
$\varphi\le\varphi_M$, and negative for $90^\circ\ge\varphi>\varphi_M$.
For larger angles, when the DDI is predominantly attractive, close to $\varphi \approx 90^{\circ}$, the system can 
become unstable, requiring enough repulsive contact interactions and suitable strong pancake-shaped trap. 
In view of these requirements, in order to stabilize the binary dipolar rotating system also in this extreme dipolar 
condition, along this work we have assumed a quite strong aspect ratio $\lambda=20$ for
the trap, with the intra-species contact interactions ($a_{ii}=50a_0$) enough large and equal for both atomic elements.

In the next subsection, we provide more details on the specific parametrization and numerical approach
considered in our study.

\subsection{Parameters and numerical approach}
The magnetic dipole moments of the considered dipolar atoms given in terms of the Bohr magneton $\mu_B$, are the following:
$\mu=10\mu_B$ for $^{162,164}$Dy, $\mu=7\mu_B$ for $^{168}$Er, and $\mu=1\mu_B$ for $^{87}$Rb.
In particular, considering the corresponding dipole moments, the strengths of the DDI are given as 
$a_{ij}^{(d)}=131\,a_{0}$ ($i,j=1,2$), for the $^{164}$Dy-$^{162}$Dy mixture; and
$a_{11}^{(d)}=66\,a_{0} $, $a_{22}^{(d)}=131\,a_{0}$ and $a_{12}^{(d)}=a_{21}^{(d)}=94\,a_{0} $, 
for the $^{168}$Er-$^{164}$Dy mixture. As the magnetic moment of $^{87}$Rb is negligible in comparison with
the $^{164}$Dy, we assume the binary mixture $^{164}$Dy-$^{87}$Rb as being single-species dipolar, such that
$a_{22}^{(d)}=a_{12}^{(d)}\approx 0$.
To explore the anisotropic effect of the DDI, we vary 
 the polarization angle $\varphi$, which influence the inter-, and intra-species dipolar interactions. 
Initially the dipoles are considered to be polarized along 
the $z-$axis with the polarization angle $\varphi=0^\circ$, which provides repulsive DDI.  
 Since, we do not have a separate control to tune the inter-species dipolar interaction,    
tuning the polarization angle of dipoles $\varphi>0$ starts to provide attractive dipolar 
 interaction in all the components of inter- and intra-species dipolar interaction. 
Apart from dipolar parameters, other intrinsic properties of binary systems are related to the two-body contact 
interactions, which in principle can be varied by using Feshbach techniques~\cite{1998inouye}.
By keeping our study in stable systems, all the two-body scattering lengths are assumed to be positive, 
with the intra-species ones  being identical and given by $a_{11}=a_{22}=50a_0$ ($a_0$ is the Bohr radius), 
with the inter-species one, $a_{12}$, obtained from the ratio $\delta=a_{12}/a_{11}$. 
Relevant for the vortex-patterns that we are studying, we allow $\delta$ to be changed in a region of interest 
in the parameter space, from smaller to larger values. The  $\delta$ is considered as follows, for the miscible 
mixture $\delta=0.75$, intermediate level where one may find partially miscible at $\delta=1.0$ 
and for completely immiscible mixtures $\delta=1.1$, 1.25 and 1.45.  The rotation frequency smaller than 
$\Omega<0.4$ may not enough to observe vortex lattice structures. From the previous study and numerical 
test, we choose the rotation frequency $\Omega = 0.6$, which is enough to produce the vortex lattice structures 
within the  parameters that we considered in the present study. 
For the harmonic trap potential, a strong pancake-shaped trap in the $(x,y)-$plane is applied, 
with an aspect ratio $\lambda =20$, with the atom numbers fixed for both species at 
$N_{1}=N_{2}=10^{4}$. These choices are appropriate for experimental realistic settings due to 
stability requirements.  
Once defined the transversal trap frequency for the first species as given by $\omega_{1} = 2\pi\times 60 {\rm s}^{-1}$, 
the trap frequencies for the second species can be found from our assumption on the angular frequencies such that 
$m_2\omega_2^2\simeq m_1\omega_{1}^2$, implying that the trap frequencies are about the same for both species 
in the cases of  $^{164}$Dy-$^{162}$Dy and $^{168}$Er-$^{164}$Dy binary mixtures; whereas for the $^{164}$Dy-$^{87}$Rb
the trap frequency of the rubidium is $\omega_{2} \approx 2\pi\times 82 {\rm s}^{-1}$.

In our analysis of coupled binary mixtures, a relevant property that affects strongly the vortex patterns is 
the miscibility of the system, which varies according to the inter- and intra-species interactions 
(contact and/or dipolar in the present case). In order to estimate that, we use a parameter defined and studied 
in Ref.~\cite{2017jpco}, given by 
\begin{equation}\eta\equiv \int |\psi_1| |\psi_2| dxdy ,\label{eta}\end{equation} 
implying in $\eta = 1$ for a complete miscible system, with $\eta=0$ for a complete immiscible one.

Related to the numerical procedure to solve Eq.~(\ref{2d-2c}), as in Ref.~\cite{2017jpco}, we apply the 
split-step Crank-Nicolson method~\cite{CPC2,Gammal2006}, combined with a standard method for evaluating DDI integrals in 
the momentum space (see details in Ref.~\cite{Kumar2017} and references therein). 
In order to search for stable solutions, in the present work the numerical simulations were carried out in imaginary time on a grid 
with a maximum of 528 points in each $x$ and $y$ directions $x$ and $y$, with spatial and time steps $\Delta x=\Delta y=0.05$ 
and $\Delta t=0.0005$, respectively. 
Both component wave functions are renormalized to one at each time step.
In order to obtain stationary vortex states, we solve the Eq.~(\ref{2d-2c}) with different initial conditions. 
In view of previous tests on the initial suitable conditions, we use combination of angular harmonics 
followed by convergence tests of the solutions for the given inputs, within a procedure also discussed in 
Ref.~\cite{Kumar2017}.

\section{Results on binary dipolar mixtures}
Our main results are presented in this section, which are verified by tuning the DDI from repulsive to attractive,
leading to an almost complete spatial phase separation of rotating binary mixtures at some critical orientation
of the dipoles, when the DDI becomes attractive. 
We consider three dipolar coupled BEC mixtures, which have different natural characteristics in respect to 
their miscibility properties, as well as different vortex-pattern structures when the binary system is under
rotation, as verified in previous studies~\cite{2017jpco,Kumar2017}.
First, we consider a symmetric-dipolar coupled system, $^{164}$Dy-$^{162}$Dy, where both magnetic dipoles 
have the same value,  which is more miscible when both species have the same polarization, $\varphi=0$. Next, we
consider an asymmetric-dipolar mixture, the $^{168}$Er-$^{164}$Dy, which is less miscible, having both species 
with large but different magnetic dipoles. Finally, in order to compare with the previous two cases, we also examine the 
$^{164}$Dy-$^{87}$Rb case, considering it as a single-dipolar binary mixture, as one of the species have negligible 
dipole moment.

In all the cases that we are presenting, as explained in the previous section, some of the parameters are fixed to 
approximate realistic values considered in experiments, as well as fixed by favorable stability conditions.
With the parameters as number of atoms for each species, trap aspect ratio, intra-species scattering lengths and
rotation frequency settled, our study is concentrated in the main focus of our work, which is to explore the
behavior of a strong dipolar coupled system, by tuning the polarization angle of the dipoles  (varied from 
zero to 90$^\circ$) together with variations of the inter- to intra-species two-body contact interactions $\delta$.
In view of the miscibility properties of the binary dipolar mixtures, the parameter $\delta$
determines the initial vortex patterns (when $\varphi=0$ and the DDI is repulsive). By increasing $\varphi$, the DDI
can be modified to be attractive, which happens for $\varphi>54.7^\circ$.

\begin{figure}[tbp]
\includegraphics[width=0.5\textwidth]{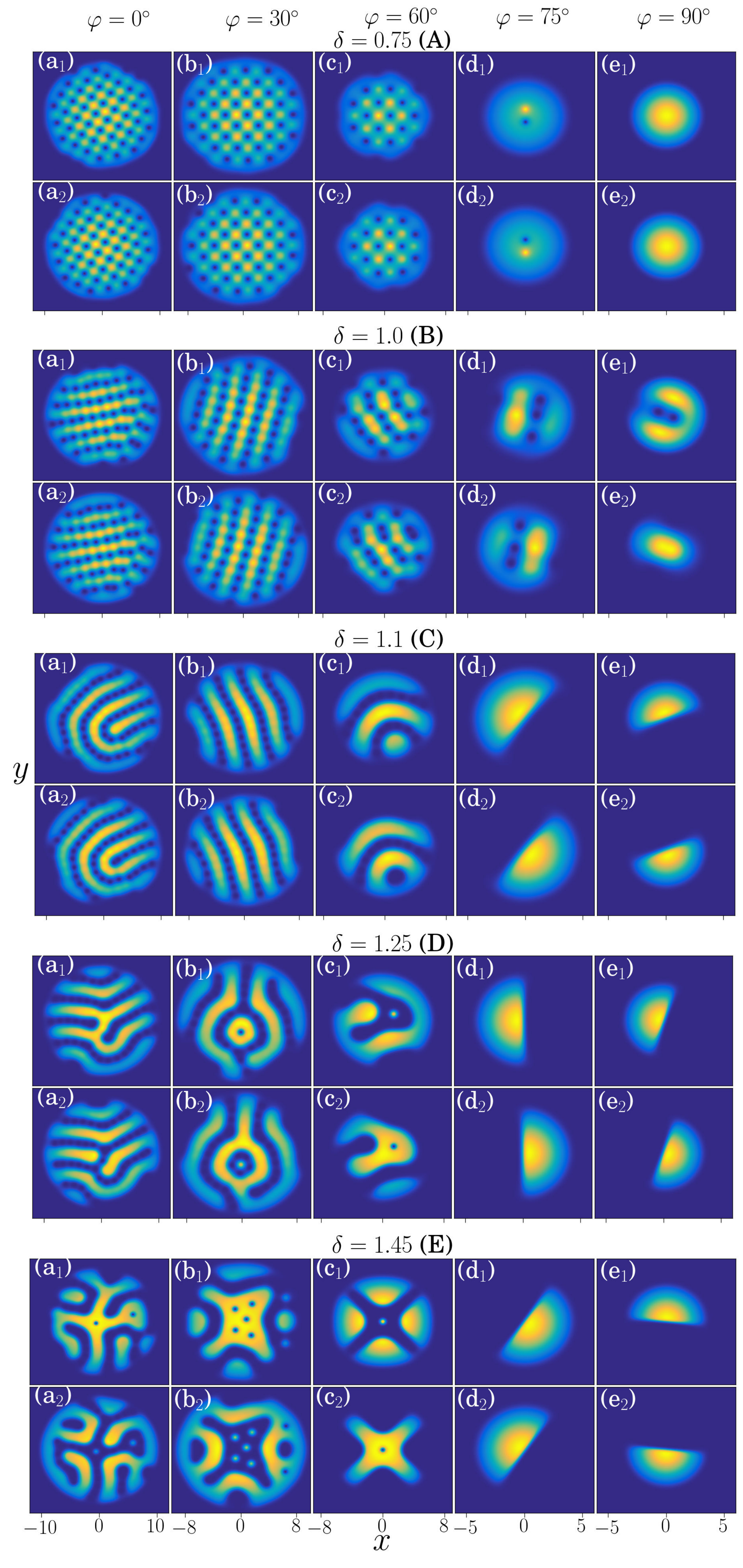}
\caption
{(Color on line) The 2D densities $|\psi_{j=1,2}|^2$ are shown for the $^{164}$Dy-$^{162}$Dy dipolar mixture ($j=1$ is the $^{164}$Dy, 
with $j=2$ the $^{162}$Dy), by tuning the polarization $\varphi$ from 0$^\circ$ [(a$_{j}$)] to 90$^\circ$ [(e$_{j}$)], with 
$\delta$ varying from 0.75 [set (A)] to 1.45 [set (E)]. 
All panels have square $(x,y)$ dimensions, indicated by the $x-$labels.
The $(x,y)$ and $|\psi_{j}|^2$ are dimensionless, with $l_\perp=1\mu$m being the space unit.
The density levels vary from 0 (darker) to a limit ranging $0.009 \sim 0.12$ (lighter), fixed by their respective 
normalization to one. 
}
\label{fig02}
\end{figure}

\begin{figure}[tbp]
\begin{center}
\includegraphics[width=0.45\textwidth]{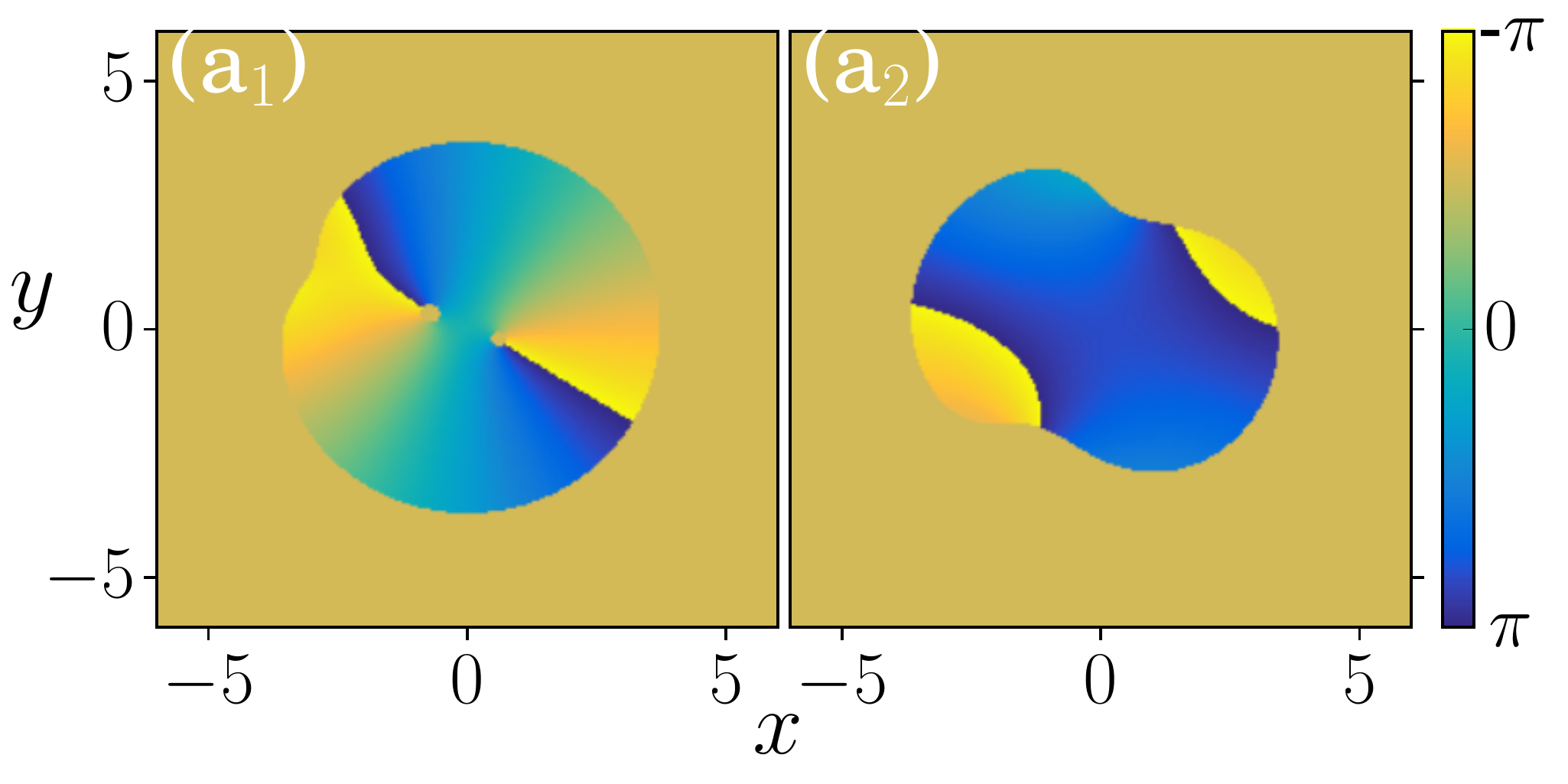}
\end{center}
\caption{(Color on line) Phase diagrams  (a$_{j=1,2}$) for the densities $|\psi_{j}|^2$ of the double-core vortices, corresponding 
to panels (e$_j$) of Fig.~\ref{fig02}(B), when the DDI is purely attractive ($\varphi=90^\circ$).
The $(x,y)$ and $|\psi_{j}|^2$ are dimensionless, with $l_\perp=1\mu$m being the space unit.
}
\label{fig03}
\end{figure}

\begin{figure}[tbp]
\begin{center}
\includegraphics[width=0.45\textwidth]{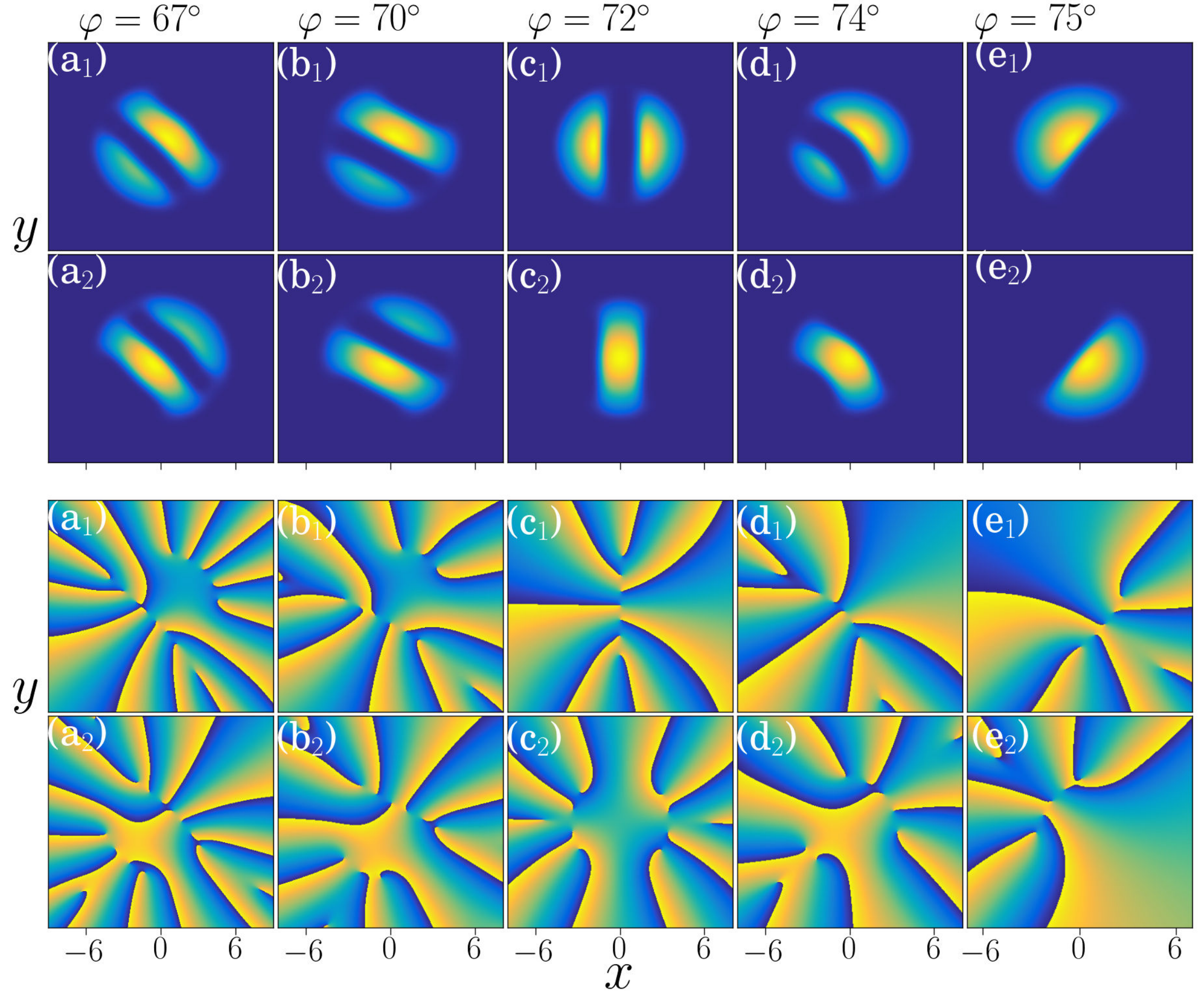}
\end{center}
\caption{(Color on line) 2D densities $|\psi_{j=1,2}|^2$ are shown in the upper two sequence of panels, 
for the $^{164}$Dy-$^{162}$Dy mixture, as in Fig.~\ref{fig02}(C) ($\delta=1.1$), but considering the  
tuning-polarization angle from 67$^\circ$ [(a$_{j}$)] to 75$^\circ$ [(e$_{j}$)]. 
The corresponding phase diagrams, using the same respective labels (a$_{j}$) for 67$^\circ$ to (e$_{j}$) for 75$^\circ$, 
are given in the lower two sequence of panels, with hidden vortices being confirmed.
The $(x,y)$ and $|\psi_{j}|^2$ are dimensionless, with $l_\perp=1\mu$m being the space unit.
} 
\label{fig04}
\end{figure}

\begin{figure}[tbp]
\begin{center}
\includegraphics[width=0.45\textwidth]{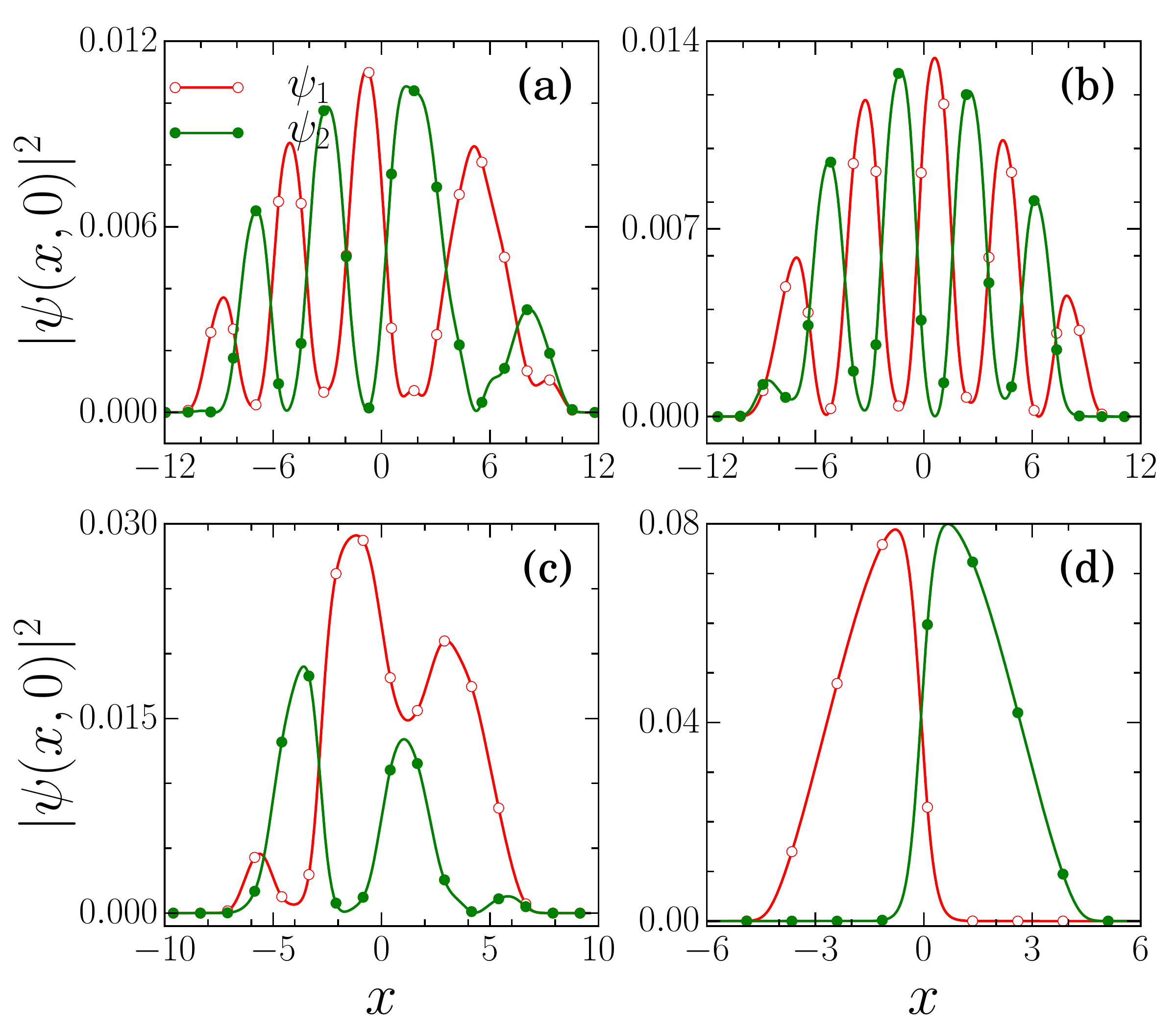}
\end{center}
\caption{(Color on line)
Densities, given in 1D for $y=0$, are shown for the binary mixture $^{164}$Dy-$^{162}$Dy, 
corresponding to $\delta=1.1$, Fig.~\ref{fig02}(C), for the polarization angles $\varphi\,=\,0^\circ$ (a), 
$30^\circ$ (b), $60^\circ$ (c), and $75^\circ$ (d). 
In particular, the panel  (d) shows the mixture reaching an almost-complete spatial separation of the densities. 
The $(x,y)$ and $|\psi_{j}|^2$ are dimensionless, with $l_\perp=1\mu$m being the space unit.}
\label{fig05}
\end{figure}

 \begin{figure}[tbp]
\begin{center}
\includegraphics[width=0.45\textwidth]{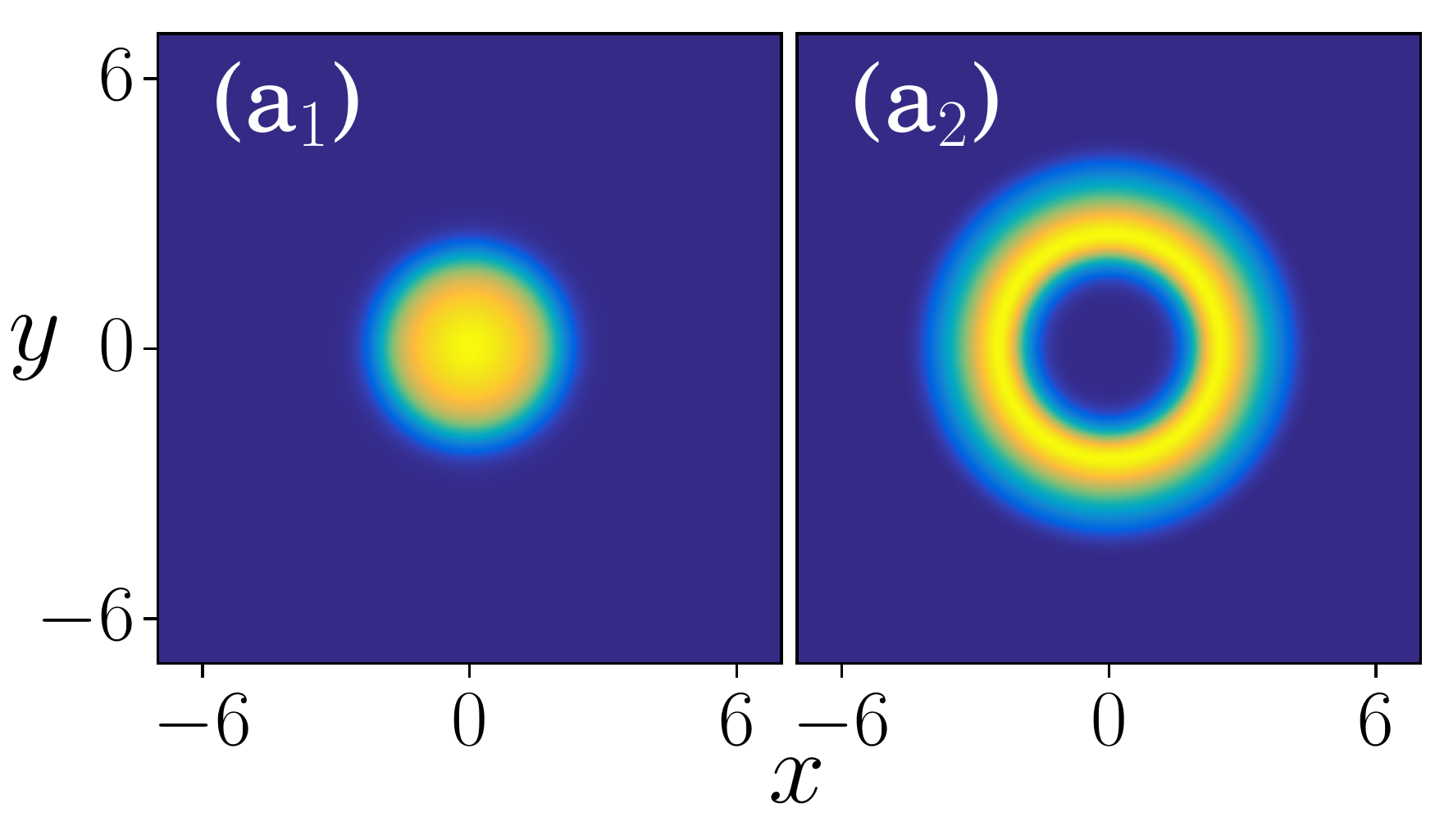}
\end{center}
\caption{(Color on line) Nonrotating ($\Omega=0$) condensate densities for the dipolar mixture $^{164}$Dy-$^{162}$Dy, considering
 $\delta=1.1$ and $\varphi\,=\,75^\circ$, corresponding to panels (d$_{j}$) of Fig.~\ref{fig02}(C).  
 The $(x,y)$ and densities are dimensionless, with $l_\perp=1\mu$m being the space unit.
}
\label{fig06}
\end{figure} 
\begin{figure}[tbp]
\begin{center}
\includegraphics[width=0.45\textwidth]{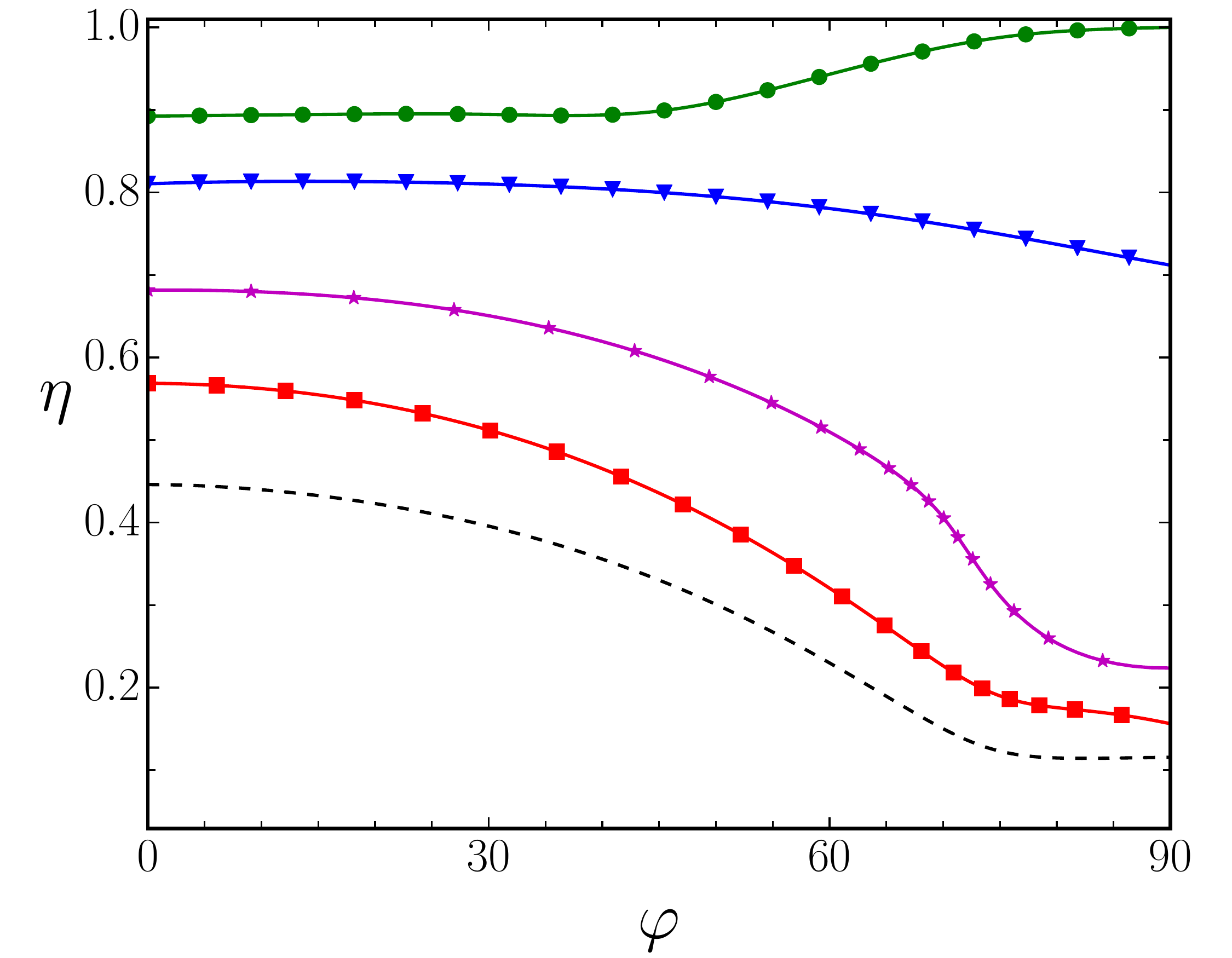}
\end{center}
\caption{ (Color on line) The miscibility parameter ($\eta$), defined in Eq.~(\ref{eta}), is shown as function of the polarization angle $\varphi$,
for the five cases of inter- to intra-species contact interactions presented in Fig.~\ref{fig02}.
For $\delta=0.75$, we have the green-solid line with circles; for $\delta=1.0$, the 
blue-solid line with triangles; for $\delta=1.1$, the purple-solid line with stars;
for $\delta=1.25$, the red-solid line with squares; and for $\delta=1.45$ the dashed line.
}
\label{fig07}
\end{figure}
\begin{figure}[tbp]
\begin{center}
\includegraphics[width=0.45\textwidth]{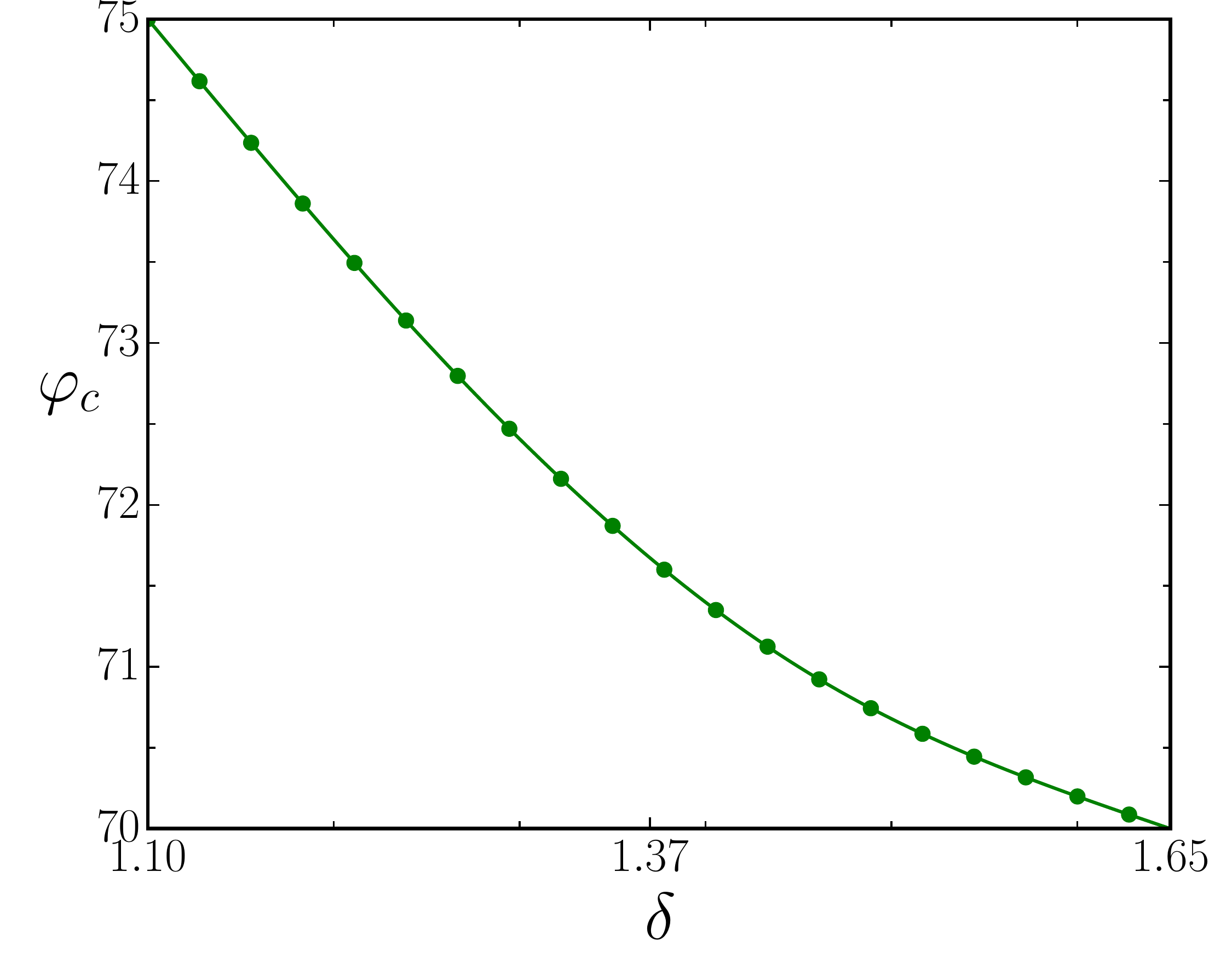}
\end{center}
\caption{Critical angle $\varphi_c$ for the almost-complete spatial  separation of the mixture
$^{164}$Dy-$^{162}$Dy, considering the relevant interval region of $\delta$ that we are studying.}
\label{fig08}
\end{figure}

\subsection{The symmetric-dipolar $^{164}$Dy-$^{162}$Dy binary mixture}

Within our aim to analyze the anisotropic property of the dipolar interactions by tuning the dipoles from 
$\varphi=0^\circ$ to $90^\circ$, we first consider the nearly symmetric mixture $^{164}$Dy-$^{162}$Dy, 
where both dipolar species have the same magnetic dipole moments, with the coupled system being more 
miscible than the other two cases we are studying; at least when the DDI is repulsive. 
As verified in Ref.~\cite{Kumar2017},  this dipolar mixture can show triangular, squared, striped, and domain wall 
vortex-lattice structures regarding to the ratio between inter- and intra-species contact interaction $\delta$ and 
rotation frequency $\Omega$. The rotation in the present study is fixed to 0.6$\omega_\perp$, guided from a 
previous study in Ref.~\cite{Kumar2017}, as found it is appropriate (not too large or small) to study other properties 
of binary dipolar systems trapped in 2D pancake-type geometries, such as the behaviors due to changes in the 
contact and dipolar interactions. 
In the absence of contact interaction, the inter-, and intra-species dipolar interactions are balanced and behaves like 
the case having $\delta=1$.
So, the main characteristics of this dipolar symmetric system is that, being more miscible it is expected to present 
striped-vortex patterns as $\delta$ is increased, when the dipoles are both polarized in the same direction, which 
are clearly verified in the panels (a$_j$) of Fig.~\ref{fig02}(B). The density patterns change quite significantly by 
increasing $\delta$, particularly when the coupled system becomes attractive.
 
From the results shown in Fig.~\ref{fig02}, where we have a few specific sets for fixed values of $\delta$ from 
0.75 till 1.45, one can extract some general characteristics.
First we notice that the sizes of the condensates (each species $j=1,2$) remain approximately the same as we 
increase $\delta$, for some given polarization angle $\varphi$. As, in this case, the dipolar parameters are also fixed, together 
with the intra-species contact interactions ($a_{jj}=50a_0$), we are verifying that the relevant role of the inter-species 
contact interaction ($a_{12}= a_{jj} \delta$) is mainly in the vortex-pattern structures and miscibility, with practically no 
interference on the sizes of the condensates. Only for fixed angles $\varphi\ge 75^\circ$ (when the DDI is more attractive) 
we observe some small effect of $\delta$ in the radius, which slightly increase for larger values of $\delta$; an effect which 
will be clarified  as due to the formation of vortices in the coupled system as $\delta\ge 1$, resulting in some extra repulsive
internal forces opposing the attractive DDI.

For the DDI, we should first observe that  the polarization angle $\varphi$ is the only parameter affecting both inter- and 
intra-species dipole interactions in the same way.  
However, as already observed in the case of contact interactions, the main responsible for the sizes of the density 
distributions are the intra-species forces. As the corresponding contact interactions are kept fixed, the sizes of the
density distributions  are reduced by increasing the attractive DDI, $\varphi$ approaching 90$^\circ$.
The condensates shrink due to more intra-species dipolar attraction.
The DDI is averaging exactly to zero at the particular case for $\varphi=\varphi_M=54.7^\circ$, where  $\varphi_M$
is the so-called ``magic-angle".  So, at this angle, no effect is expected from the DDI, such that the observed vortex-pattern 
structures should only due to the rotation angle, contact interactions and corresponding mass differences of  the coupled 
mixtures. 

The most striking result we found in this symmetric-dipolar case occurs due to the mass-symmetry breaking, represented
by the set with $\delta=1$, in Fig.~\ref{fig02}(C).   
By considering $m_2=m_1$ in Eqs.~(\ref{2d-2c}) and (\ref{par}), the coupled two-component equation reduces to a 
single-component one, implying that the results presented in the upper and lower panels of Fig.~\ref{fig02}(C) should be 
identical. This is confirmed by our numerical approach in solving the coupled system  Eq.~(\ref{2d-2c}).
We should noticed that, the mass asymmetry we have in this case, $\delta_M=1/81$, is quite small and appears only in 
two terms of the dimensionless Eq.~(\ref{2d-2c}). 
The results shown in Fig.~\ref{fig02}, by considering a fixed $\delta>1$, are reflecting the relevant changes in the vortex 
patterns when we increase
this polarization angle near $\varphi_M$ (from 30$^o$, when it is repulsive, to 60$^o$, when it is attractive). 
By decreasing the repulsion of the dipoles, the vortices merge together providing striped structures in the 
coupled densities, which happen because the coupled system is less miscible when the inter-species is larger than 
the intra-species interaction ($\delta>1$). 

In an analysis provided in Ref.~\cite{2009-kasamatsu-pra}, when considering that the only nonlinear repulsive forces in 
a rotating coupled system are due to contact interactions, the structure and boundary domains of stripes
are discussed by using an approximate Thomas-Fermi profile. However, the analysis was restricted to the symmetric
case with $\delta=1$, when all the masses and frequencies and number of atoms are identical for both species. 
This is further verified in the same reference that is quite approximate for $\delta>1$.
In the present case, the corresponding behavior can be directly followed from results shown in Fig.~\ref{fig02}, 
by considering in particular the columns (a$_i$) to (c$_i$), for fixed repulsive DDI, when $\varphi\lessapprox 60^\circ$.
As shown, the stripes become larger by increasing the inter-species interaction $\delta$, 
changing to serpentine-like and domain-wall patterns. This implies that to estimate how the boundaries of the
stripe increase one cannot rely in an approximation in which there is no asymmetry in the contact interactions 
(when all the other parameters for the species are identical), 
as this will correspond to a single-species interaction.

Correspondingly, one could try to extend the analysis performed in Ref.~\cite{2009-kasamatsu-pra} to the cases that we 
have $\delta$ fixed, changing the DDI by tuning the angle $\varphi$.
As shown in Fig.~\ref{fig02}(C), for example,  the stripes are becoming wider when the DDI is more attractive; 
presenting a behavior not similar as the observed for the contact interactions.
The problem is that both dipolar and contact interactions are not being varied in the same way.
When we vary the DDI by tuning the dipoles, all the inter- and intra-species are
being changed in the same way, such that the ratio between inter- and intra-species dipolar interaction remains fixed,
whereas when changing $\delta$ only the inter-species interactions are being modified. 
So, for the DDI, the dominant behavior which is being verified is due to the increasing intra-species dipolar attraction when 
moving $\varphi$ to larger angles. The condensates start to shrink, reducing the space of the overlap between the two 
species. From one side, by increasing $\delta$, as shown before the binary system becomes less miscible; and from 
the other side, the coupled condensate are shrinking by increasing the DDI. Together, the interplay of these two effects 
will result in an almost complete space separation between the densities, as observed in Fig.~\ref{fig02} for 
$\varphi\ge 75^\circ$ with $\delta>1$.  Considering that this mixture is dipolar symmetric, the tendency of both
species in occupying the same space (not possible due to the miscibility properties) leads to the
observed final half-space occupancy for each one of the two components.
A critical polarization angle can be defined in this case, which is happening near $\varphi=75^\circ$. In order to be more
precise, we define such critical angle  $\varphi_c$ as being given by the condition that  the densities of both species are 
well  defining two maxima, one for each species. 
This kind of separation of the densities, 
which we call ``angular spatial separation", occurs for symmetric dipolar mixtures in the immiscible phase.  As it will be
shown, for the asymmetric cases in their immiscible phase, the spatial separation is radial instead of angular.

The miscibility of the coupled mixture is estimated by using the parameter defined in Eq.~(\ref{eta}), which 
indicates the overlap between the two densities. The results presented in Fig.~\ref{fig02} are shown to be compatible 
with the ones obtained for non-dipolar systems, given in Ref.~\cite{2017jpco} (See figure 9 of this reference).
from where we can extract that, for $\delta=0.75$ we have an almost complete miscible system ($\eta=1$), whereas for 
$\delta\approx 1.5$ the system becomes almost immiscible ($\eta\approx 0.35$). 

In the next, the results given in Fig.~\ref{fig02} are being analyzed by considering the sets with different values of $\delta$. 
Our analysis is complemented with the next Figs.~\ref{fig03}-\ref{fig08}.
Among the sets with different $\delta$ values, the more miscible binary mixture occurs for $\delta=0.75$, given in 
Fig.~\ref{fig02}(A). With this value for $\delta$ and with the rotation frequency $\Omega=0.6$, we can avoid 
the collapse of the mixture as the polarization angle is tuned to 90$^\circ$ (more attractive case).
This mixture produces squared-lattice structure at $\varphi=0$. Some distortion are observed near 
$\varphi\,=\,45^\circ$, but the square-lattice patterns are sustained up to $\varphi\,\approx\,60^\circ$, as shown in the 
panels (a$_{j}$)-(c$_{j}$) of Fig.~\ref{fig02}(A). 
As $\varphi$ increases, the number of vortices gradually decreases, with a single vortex being shown at 
$\varphi=75^\circ$, and disappearing at $\varphi=90^\circ$, when the rotation $\Omega=0.6$ is verified not to
be enough to form vortices. 
So, the miscibility of the coupled system becomes almost complete ($\eta\approx 1$)
at $\varphi\sim 90^\circ$ due to the dominance of the attractive dipolar forces. 

A critical rotation frequency ($\Omega_c$) exists to produce vortices, which can be approximately obtained by using the 
Feynman rule~\cite{2009Fetter}. According to this rule, derived within a Thomas-Fermi approximation for a single condensate, the
total number of vortices (generated by a given rotation frequency) should correspond to $N_v=2\Omega \langle x^2\rangle$
(considering our dimensionless units in a quasi-2D spherically symmetric condensate).
This rule is only approximately verified in our case when considering a complete repulsive system ($\varphi=0$),
as counting the visible vortices for each component. As we move to attractive systems (higher polarization angles), 
less number of vortices are shown than the predicted one.
However, as already verified for pure dipolar condensates in double-well potential,  in Refs.~\cite{2017-Sabari,2014-Mithum}, this 
rule is satisfied when also counting the possible hidden vortices, which can be identified by the phase singularity distributions.

By increasing $\delta$, we show in the set (B) of  Fig.~\ref{fig02}(B) ($\delta=1$) the effect of tuning the dipoles, starting with
striped vortices ($\varphi\,=\,0^\circ$ and 30$^\circ$), as verified in the panels (a$_{j}$) and (b$_{j}$). 
By increasing $\varphi$, striped-shaped vortices are maintained up to $\varphi\,=\,40^\circ$. When the polarization angle 
$\varphi\,>\,40^\circ$, the number of vortices are not enough to form stripes, with the appearance of double-core vortices. 
Further increasing of $\varphi$ up to 90$^\circ$, only one double-core vortex (for each species $j=1,2$) can be observed.
The corresponding phase diagrams of the double-core vortices shown in (e$_{j}$) of Fig.~\ref{fig02}(B)
are presented in (a$_{j}$) of Fig.~\ref{fig03}. 
As we increase even more $\delta$ to 1.1, 1.25, and 1.45, we enter in the immiscible region of the  
$^{164}$Dy-$^{162}$Dy binary  mixture, which is shown in the sets (C), (D) and (E), respectively,
of Fig.~\ref{fig02}. In the repulsive cases, shown in the panels (a$_j$) and (b$_j$), we observe
striped-shaped vortices when $\delta=1.1$, changing to domain-wall structures for $\delta=$1.25 and 1.45.
For the given $\delta$ values, this mixture is space separated radially between $\varphi=0^\circ$ and 
some critical orientation angle ($\varphi_c$). In this kind of spatial  separation, the component
with heavy mass (here, component $j=1$, $^{164}$Dy) is located between the lighter one ($^{162}$Dy, $j=2$).  
The lighter one is proliferated around the heavy component.  
The striped-vortex patterns observed for $\delta=1.1$ at $\varphi=0$ continued up to $\varphi$ close to 60$^\circ$, as 
seen in (c$_j$) of Fig.~\ref{fig02}(C). Further, for $\varphi\,\ge 75^\circ$, we observe an almost complete spatial separation of the
mixture.  

In order to understand how the spatial  separation is occurring in this symmetric dipolar mixture, we consider 
Fig.~\ref{fig02}(C), when $\delta=1.1$, for polarization angles close to the critical value where an almost complete
spatial  separation is observed. For that, we provide in Fig.~\ref{fig04} a set of panels with the densities and the corresponding 
phase diagrams for the vortex states, by considering the angle $\varphi$ varying from 67$^\circ$ to 75$^\circ$. 
With these panels, we show how we define the critical angles. As we increase $\varphi$ in this interval, we observe 
that the initial two maxima verified for $\varphi=67^\circ$, for each component, reduce to just one maxima when 
$\varphi=\varphi_c=75^\circ$, as in the panels  (d$_j$) of Fig.~\ref{fig02}(C) (For  $\varphi=74^\circ$ we can still observe
two maxima in the densities of one of the components).
Interesting enough is the observation of hidden vortices where the component densities are practically absent (at the minimum). 
These hidden vortices provide enough repulsion for an almost complete spatial  separation of the mixture. 
The densities for the coupled mixture shown in Fig.~\ref{fig02}(C), for $\delta=1.1$, is further analyzed within one-dimensional 
plots given in four panels of Fig.~\ref{fig05}, for $y=0$, with $\varphi$ going from 0 to $75^\circ$.
The almost complete spatial  separation of the mixture can be observed at the panel (d) of Fig.~\ref{fig05}. 

The sets D and E of Fig.~\ref{fig02}, respectively for  $\delta=$ 1.25, and 1.45, are also showing the same behavior of almost
complete spatial  separation in the attractive part of the dipolar interaction, when the polarization angle $\varphi\ge 75^\circ$. 
The domain-wall vortex patterns verified when $\varphi=$ 0 and 30$^\circ$ start changing when the interaction becomes attractive, 
at $\varphi=60^\circ$, until we have the almost complete spatial  separation at some critical angle $\varphi_c$. 
Before that, for intermediate angles $\varphi<\varphi_c$, it is possible to observe rotating droplet states, as we can verify 
particularly in the panels (a$_{j}$)-(c$_{j}$) of Fig.~\ref{fig02}(E).
 These kind of rotating droplet structures have already been observed in Ref.~\cite{2003-Kasamatsu}, when considering particular
 values of  $\delta$ and rotation angles $\Omega$.
 
Due to the immiscibility, rotating droplet density  peaks are formed near to the surface of the BEC in the first component;
being located near the middle of the condensate for the second component. In the structures of the 
rotating droplets vortex lattice, we observe about vorticity two in the small-sized ones, and about four in the 
large-sized ones. 
They are similar as the double-core structures, where rotating droplets in any one of the components are formed by multiple vortices with 
the same circulation, with vortices in any of those components being surrounded by multiple vortices. 
 The rotating droplet shaped density separations can be observed only within particular parameter regimes of $\delta-\Omega$,
 when the number of vortices is enough to produce droplets. 
 For the value of $\Omega=0.6$ we are considering, this condition with rotating droplets being produced is exemplified in the panels 
 (b$_j$) of Fig.~\ref{fig02}(E) (when $\delta=$1.45 and $\varphi=30^\circ$). 
 By increasing the polarization angle $\varphi$, as exemplified in the panels (c$_j$) of Fig.~\ref{fig02} (when $\delta=1.45$), 
 the number of vortices are reduced due to the attractive contributions in the DDI, such that rotating droplets are suppressed.  

The effect of rotation, in the case we have an almost complete spatial separation shown for $\delta=1.1$ and $\varphi=75^\circ$
in panels (d$_j$) of Fig.~\ref{fig02}(C), can be verified by switch-off the rotation to $\Omega=0$. The results, presented in the panels
(a$_{j=1,2}$) of Fig.~\ref{fig06}, show that  in the non-rotating regime the spatial  separation is radial, due to the dominant inter-species 
contact interaction. Such behavior confirms that the observed spatial  separation happens in rotating binary mixture due to repulsive forces 
provided by hidden vortices. 
In immiscible mixtures of rotating $^{164}$Dy-$^{162}$Dy condensates, the spatial  separation is angular, due to hidden vortices located in both 
components that are repelling each other. By increasing the polarization angle $\varphi$, the DDI becomes more
attractive, with a critical angle ($\varphi_c$) possible to be identified for an almost complete spatial  separation of the mixture. 
 
 As the dipole polarizations can change the DDI from repulsive to attractive, it is expected that it will affect the miscibility
 $\eta$ of a dipolar coupled system; in a way similar as $\eta$ being affected by the inter- to intra-species contact interaction 
 $\delta$.  In this regard, we should remind from Ref.~\cite{Kumar2017} that, by increasing $\delta$ we provide more repulsion 
 between the inter-species, implying that the system becomes less miscible (decreasing $\eta$). This effect is clearly shown by the five 
 different curves presented in Fig.~\ref{fig07} when we assume fixed DDI. 
 By increasing the polarization angle $\varphi$ from $0^\circ$, we have the addition of a repulsive DDI effect, which is being reduced
 till $54.7^\circ$, when the DDI is zero. However, when changing $\varphi$, all the inter- and intra-species DDI interactions are varying in 
 the same way, keeping the same ratio, such that the main miscibility behavior given by $\delta$ is expected to remain.
 
 And, by further increasing $\varphi$, for $\varphi>54.7^\circ$, we have the inter-species repulsive effect due to $\delta$ being 
 diminished; at the same time, the attractive intra-species DDI is increasing, which should improve the miscibility of the two 
 species. This behavior is clearly shown for the case that $\delta<1$ (upper green-line with bullets in the Fig.~\ref{fig07}).
 In this case, we should interpret that the effect of DDI (attractive contribution, in this case) is dominant in comparison with the contact 
 interaction effect. 
 However, for $\delta>1$, the coupled system is shown to become less miscible when the polarization angle is $\varphi\ge\varphi_c$, 
 besides the attraction provided by the DDI, which is still effective as verified by the shrinking of the radius. 
 As clarified before, for the sizes of the condensates, the intra-species interactions are the relevant observables, such that we conclude
 that the dipolar ones are dominant in respect to the contact ones. 
 However, for the miscibility, the main role are provided by the inter-species interactions. 
 As the system becomes less miscible for $\varphi\ge\varphi_c$, we conclude that the repulsive 
 inter-species contact  interaction are dominant in this region where the DDI is attractive.
 The combination of both two effects, one due to repulsive contact interactions (reducing the miscibility), with the other due to attractive
 DDI (reducing the size of the condensates) result in the observed half-space spatial separation of the mixture for larger polarization
 angles.  In summary, one can interpret $\varphi_c$ as a critical angle at which the system becomes dominated by the 
 DDI, if $\varphi<\varphi_c$; or by the inter-species contact interactions, if $\varphi>\varphi_c$.
 
Our results on this study are shown in Fig.~\ref{fig07}, where we are verifying that for $\delta=0.75$ the attractive DDI is dominating, when
$\varphi\ge 54.7^\circ$, increasing the miscibility $\eta$. With $\delta=1$, the miscibility of the mixture is weakly affected when tuning the 
dipoles from repulsive to attractive, but $\eta$ starts to decrease for large polarization angles, indicating the dominance of contact interactions.
 Further, for the other three cases considered in Fig.~\ref{fig02} (with $\delta>1$), we observe the dominant effect of the contact 
 interactions in the immiscibility of the mixture (diminishing $\eta$), with $\varphi_c$ being the critical position where it occurs an almost 
 complete spatial separation of the mixture, with formation of hidden vortices. 
 
 In order to estimate the critical angle for the spatial  separation, $\varphi_c$ is obtained as a function of $\delta$ in Fig.~\ref{fig08}, 
 where we observe that $\varphi_c$ decreases from $75^\circ$ to $70^\circ$ with $\delta$ increasing, respectively, from 1.1 to 1.65.
 By increasing $\delta$, the system becomes less miscible, with the spatial  separation being reached for smaller values of $\varphi$. 
Therefore, the complete angular spatial  separation is obtained as an accumulative effect of inter-species contact interactions 
together with repulsive forces due to hidden vortices, in case of rotating coupled condensates. 
In the non-rotating case, the spatial  separation is radial, as shown in Fig.~\ref{fig06}. 
 
Resuming our results for the $^{164}$Dy-$^{162}$Dy binary mixture, we have studied the anisotropic effects due to dipolar 
interactions by tuning the orientation angle of dipoles from miscible to immiscible cases. The fundamental vortex-lattice structures,
which occur for repulsive dipolar interactions ($\varphi<54.7^\circ$) remain until the attractive dipolar interactions become 
dominant in respect to the inter- to intra-species contact interactions. 

\subsection{The asymmetric-dipolar $^{168}$Er-$^{164}$Dy binary mixture}
\begin{figure}[tbp]
\includegraphics[width=0.45\textwidth]{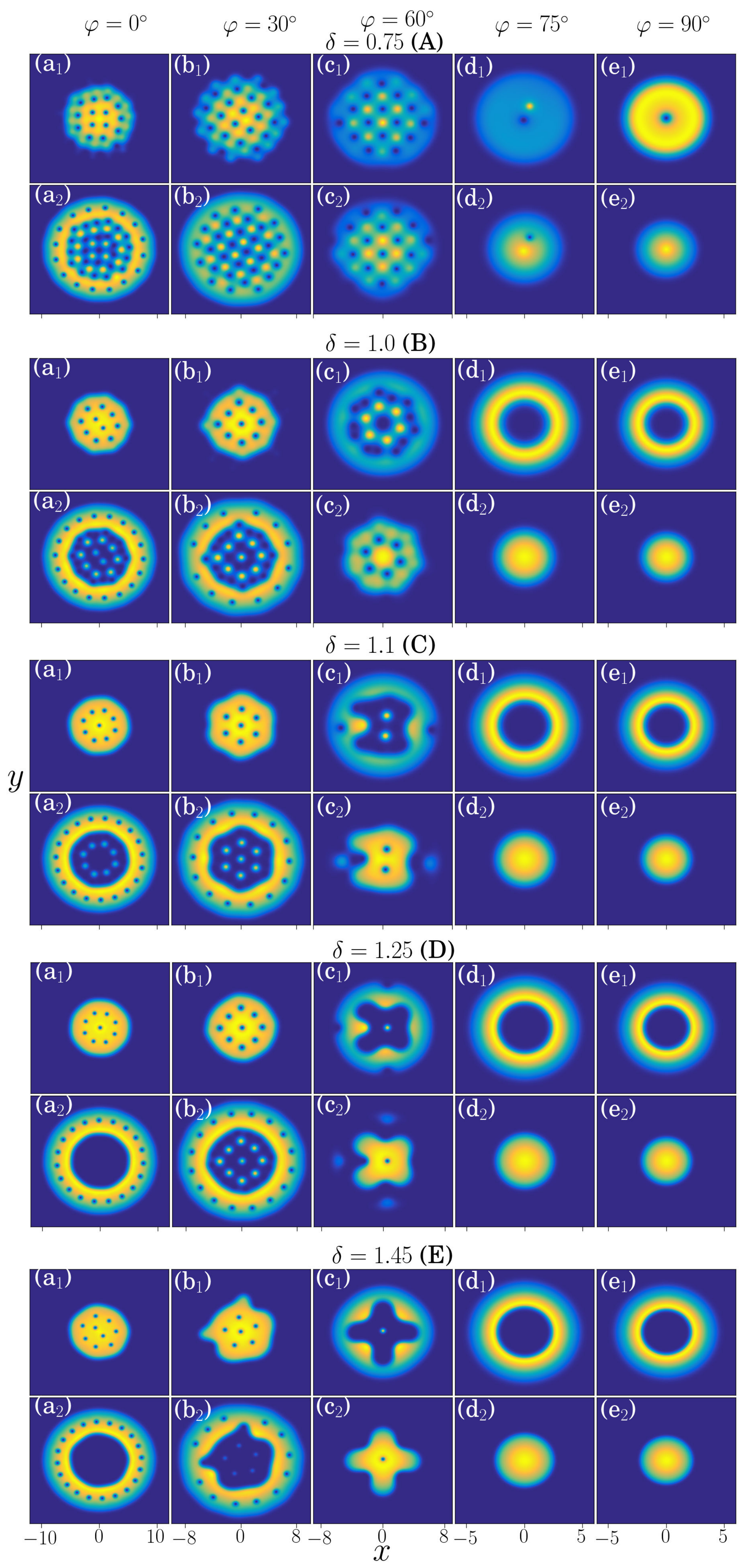}
\caption
{(Color on line) The 2D densities $|\psi_{j=1,2}|^2$ are shown for the $^{168}$Er-$^{164}$Dy dipolar mixture ($j=1$ is the $^{168}$Er, 
with $j=2$ the $^{164}$Dy), 
by tuning the polarization $\varphi$ from 0 [(a$_{j}$) panels] to 90$^\circ$ [(e$_{j}$) panels], with $\delta$ 
varying from 0.75 [set (A)] to 1.45 [set (E)]. As in Fig.~\ref{fig02},
all panels have square dimensions, such that only $x-$labels are indicated, with the corresponding
sizes decreasing from left to right.
The $(x,y)$ and $|\psi_{j}|^2$ are dimensionless, with $l_\perp$ being the space unit.
The density levels vary from 0 (darker) to a limit ranging $0.01 \sim 0.105$ (lighter), fixed by their 
respective normalization to one.
\vspace{1cm}}
\label{fig09}
\end{figure}

In this subsection, we discuss anisotropic effects of dipolar interactions for the asymmetric $^{168}$Er-$^{164}$Dy mixture. 
From previous analysis~\cite{Kumar2017}, we understand that the vortex-lattice patterns, for repulsive dipolar DDI,
mainly feature triangular, squared, and circular vortex lattice structures in this binary mixture. 
This asymmetric dipolar mixture is less miscible due to the imbalances of dipole moments and atomic masses. It is 
immiscible even when the inter- to intra-contact interactions are smaller than one ($\delta<1$), with dominant repulsive 
inter-species DDI. 
However, by tuning the dipole polarization angle we can change the DDI to be attractive, which will allow transition 
between the mixtures. 
The heavier mass $^{168}$Er component, at $\varphi=0^\circ$, is kept in the center, surrounded by the lighter mass 
one, $^{164}$Dy.  
By tuning the polarization, increasing $\varphi$, the DDI becomes attractive, 
providing the imbalanced dipolar counterpart. For example, the $^{168}$Er has less magnetic moment than the $^{164}$Dy, implying that
$^{164}$Dy will have larger attractive dipolar interaction than the $^{168}$Er. 
This will affect the corresponding radius of each component in the mixture, with the 
$^{164}$Dy component shrinking more than the $^{168}$Er component. The corresponding density distributions, observed for the
components in Fig.~\ref{fig09}, are clearly indicating this general behavior in all the sets with different values for the contact
interactions, by the exchange positions between inner and outer components, as the dipole interactions change from repulsive to attractive. 

Here, we should observe that, in the particular case when $\varphi$ is near $60^o$, we are close to the polarization
``magic angle", such that the results obtained for the panels (c$_j$) should approximately
correspond to the results presented in the  panels (c$_j$) of Fig.~\ref{fig02}. As the mass differences between both dipolar mixtures
are quite small, the observed differences in the vortex patterns are mainly due to the fact that with $\varphi=60^\circ$ we are deviating 
about 5.3$^\circ$ from $\varphi_M$, such that we have some residual dipolar effect due to this deviation.  
Next, with $\varphi>54.7^\circ$, we notice some relevant difference between the vortex-patterns obtained in Fig.~\ref{fig09}
 in relation to the ones shown in Fig.~\ref{fig02} for the symmetric dipolar case.  
 When the attractive dipolar interaction becomes dominant, with the rotation of the mixture the $^{168}$Er ($j=1$) component starts to move 
from the center to the border, with the $^{164}$Dy ($j=2)$ component moving to the center. This behavior can be explained by the fact that
the intra-species DDI is less attractive for the species with smaller magnetic moment (in this case, we have $a^{(d)}_{11}<a^{(d)}_{22}$). 
This interplay between the components can be verified in the panels (c$_j$) of Fig.~\ref{fig09}, when both components are trying
to occupy the same localization, with the DDI starting to be attractive.
For $\varphi\ge 75^\circ$ [panels (d$_j$) of Fig.\ref{fig09}], we observed that the $^{164}$Dy component is already surrounded by the 
$^{168}$Er one, implying in a radial spatial  separation. This is more pronounced for the cases that $\delta>1$, which is expected
from previous studies, as a binary mixture becomes less miscible by increasing the repulsive inter-species contact 
interactions~\cite{2017jpco,Kumar2017}.

\begin{figure}[tbp]
\begin{center}
\includegraphics[width=0.45\textwidth]{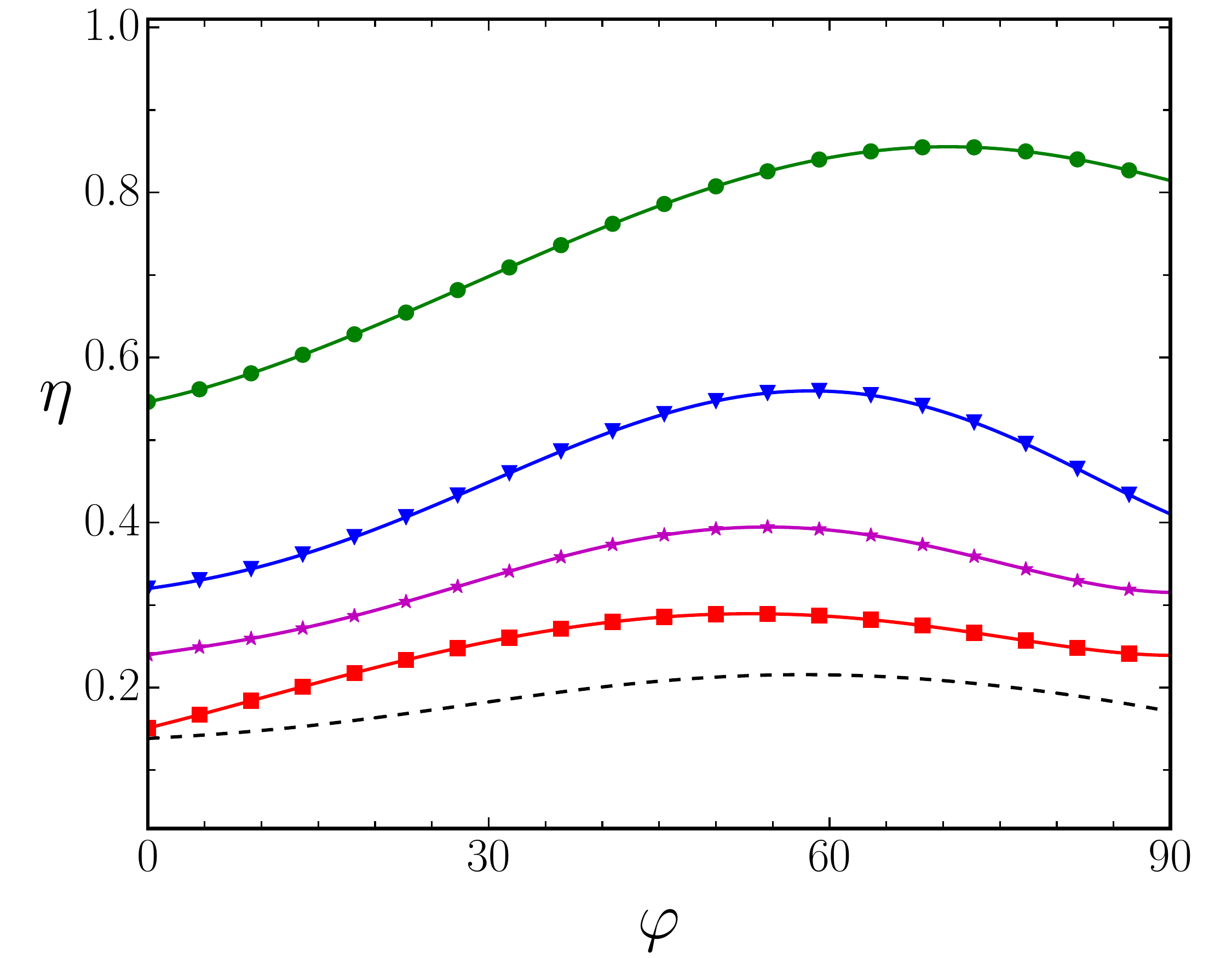}
\caption{ (Color on line) Miscibility of the dipolar mixture $^{168}$Er-$^{164}$Dy, defined in Eq.~(\ref{eta}) by the parameter $\eta$, 
is shown in respect to the polarization angle $\varphi$.
For $\delta=0.75$, we have the green-solid line with circles; for $\delta=1.0$, the 
blue-solid line with triangles; for $\delta=1.1$, the purple-solid line with stars;
for $\delta=1.25$, the red-solid line with squares; and for $\delta=1.45$ the dashed line.
}
\label{fig10}
\end{center}
\end{figure}

With Fig.~\ref{fig10} we are presenting our results for the behavior of the miscibility of this asymmetric mixture,  
$^{168}$Er-$^{164}$Dy, with respect to the changes in the polarization angle, for each of the sets of $\delta$ shown in 
Fig.~\ref{fig09}. 
In this case, for all the sets we observe a similar behavior of $\eta$ with respect to $\varphi$, with the 
coupled system becoming more miscible when increasing $\varphi$ till some critical angle where the miscibility reaches a
maximum. 
The behavior for smaller angles differs from the ones verified in Fig.~\ref{fig07} (where $\eta$ is not increasing in this
interval with $\varphi<54.7^\circ$). As Er-Dy is a mixture less symmetric than Dy-Dy, for $\varphi=0^\circ$, it is also less miscible,
such that the miscibility is increasing as the DDI becomes less repulsive. 
For larger polarization angles, we have the combined effect of attractive DDI (reducing the radius and increasing $\eta$)
together with the repulsive contact interactions till a maximum at some critical $\varphi_c$, where there is a balance
between the two combined effects. For  $\varphi\ge \varphi_c$, the contact interactions start to become dominant in relation to the
DDI. Different of the symmetric case observed in Fig.~\ref{fig07}, instead of angular spatial  separations,
here we are verifying radial spatial  separations.
As observed in Fig.~\ref{fig10}, the critical angles are in a small interval, being $\sim 70^\circ$ for $\delta=0.75$; $\sim 60^\circ$ for 
$\delta=1$; and $\sim 55^\circ$ to $\sim 60$ for $\delta=$ between 1.1 and 1.45.
Next, for $\varphi>\varphi_c$, the system starts to become less miscible with the emergence of the radial spatial  separation 
observed in Fig.~\ref{fig09}.

\begin{figure}[tbp]
\begin{center}
\includegraphics[width=0.45\textwidth]{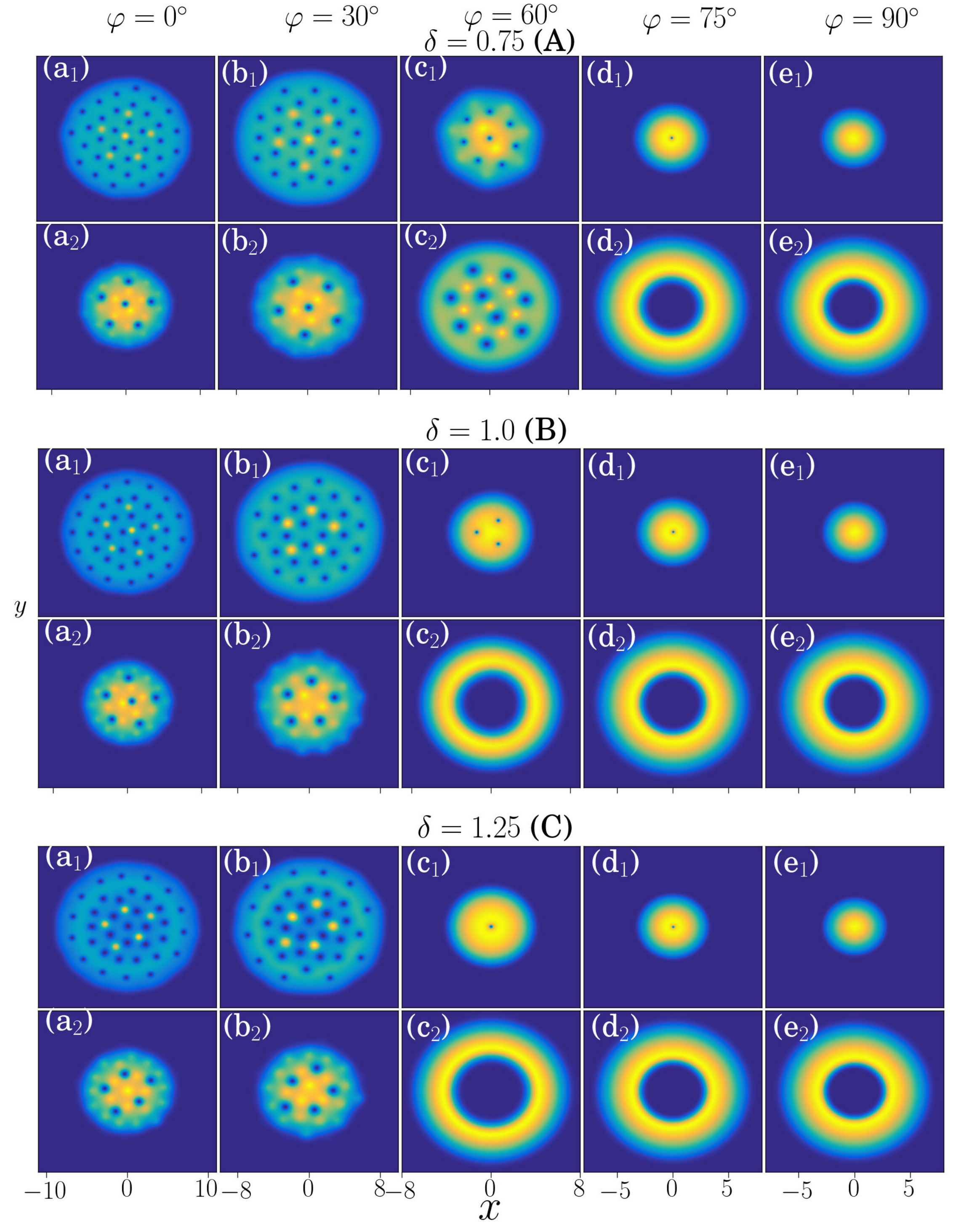}
\caption{(Color on line) 
The 2D densities $|\psi_{j=1,2}|^2$ are shown for the $^{164}$Dy-$^{87}$Rb dipolar mixture ($j=1$ is the $^{164}$Dy, 
with $j=2$ the $^{87}$Rb), 
by tuning the polarization $\varphi$ from 0 [(a$_{j}$) panels] to 90$^\circ$ [(e$_{j}$) panels], for 
$\delta=$ 0.75 [set (A)], 1.0 [set (B)] and 1.25 [set (C)]. As in Figs.~\ref{fig02} and \ref{fig09},
all panels have square dimensions, such that only $x-$labels are indicated, with corresponding
sizes decreasing from left to right.
The $(x,y)$ and $|\psi_{j}|^2$ are dimensionless, with $l_\perp=1\mu$m being the space unit.
The density levels vary from 0 (darker) to a limit ranging $0.01 \sim 0.08$ (lighter), fixed by their 
respective normalization to one.}
\label{fig11}
\end{center}
\end{figure}
\subsection{The asymmetric-dipolar $^{164}$Dy-$^{87}$Rb binary mixture}

The binary mixture $^{164}$Dy-$^{87}$Rb is more asymmetric with respect to the dipolar magnetic moments of each 
component that the other cases that we have considered, with the magnetic moment of $^{87}$Rb being almost 
negligible, such that the effect of the polarization angle is less effective, as due mostly to the $^{164}$Dy component.
In this case, the rubidium component of the mixture is more concentrated in the center, when considering the
region where the DDI is more repulsive (as shown in Fig.~\ref{fig11} for $\varphi=$ 0 and 30$^\circ$), with the
dysprosium component distributed within a larger radius. The results are quite similar for the three sets of $\delta $ that we 
are examining. When the DDI changes to an attractive one, as verified for  $\varphi\ge$ 60$^\circ$, we observe an interplay
between the distributions of the two components, with the radius of the $^{164}$Dy (component $j=1$) being strongly reduced 
in relation to the radial distribution of the  $^{87}$Rb ($j=2$). 
As in the previous discussed mixture, of erbium and dysprosium, also here the effect is clearly due to the large differences 
between the magnetic moments of both species: the DDI is more attractive for the species with larger intra-species magnetic 
moment, implying in a smaller radial distribution. In particular, we noticed that  this behavior with the corresponding radial spatial 
separation is happening even for $\delta<1$, when the DDI becomes attractive. This is reflecting the dominance of the 
attractive intra-species DDI in relation to the repulsive effects due to the intra-species contact interactions. 
Apart of that, the radial spatial  separation is similar as the one obtained for the $^{168}$Er-$^{164}$Dy mixture, when 
the DDI is attractive, with $\delta\ge 1$.

One characteristics of this mixture can be observed when $\varphi$ is near $60^o$, which is close to the polarization
``magic angle" $\varphi_M=54.7^o$  where the DDI is zero. 
The results obtained for the panels (c$_j$) are expected to correspond to the results presented in the panels (c$_j$)
of Fig.~\ref{fig09}, if the mass differences were the same. However, in the present case, the mass of the second 
species, the rubidum, is about half of the dysprosium. 
So, the results are shown a more clear separation between the species, which can only be 
explained by the mass differences, together with the larger difference in the magnetic moments.

\section{Summary and Conclusion}
\label{secIV}
In the present work we are reporting results of a study on anisotropic effects due to dipole-dipole interactions in rotating binary 
dipolar Bose-Einstein condensates, trapped in a strong pancake-type 2D geometry, obtained by full-numerical solution  of the 
corresponding coupled GP equation. 
In our approach, the rotating binary mixtures are kept stable, confined by strongly pancake-shaped trap with aspect-ratio 
$\lambda=20$ and fixed rotation frequency $0.6\omega_\perp$.
Within an investigation that can be generally extended to other kind of dipolar atomic
species, we focus our study in binary mixtures having atoms with strong magnetic dipolar properties that are under active 
investigations in cold-atom laboratories. More specifically, we consider miscible and immiscible regimes of 
$^{164}$Dy-$^{162}$Dy, $^{168}$Er-$^{164}$Dy and $^{164}$Dy-$^{87}$Rb binary mixtures, which are obtained by
changing the inter- to intra-species repulsive two-body contact interactions.
The polarization of the dipoles, given by the angle $\varphi$, is instrumental in the process of tuning from repulsive to 
attractive the dipolar interactions for the inter- and intra-atomic species.
With the dipoles of the two species polarized in the same direction, perpendicular to the direction of the dipole alignment
($\varphi=0$), the DDI is repulsive. By tuning the polarization angle $\varphi$ from zero to 90$^\circ$ the DDI 
changes from repulsive to fully attractive.  

In our study, we first notice that the space distributions of the condensed mixture of each species are mainly affected 
by the intra-species interactions (repulsive or attractive); with the inter-species interactions being responsible for the 
vortex-pattern structures, reflecting the miscibility of the mixture (by increasing $\delta$ the binary system becomes 
less miscible).
By tuning the polarization angle $\varphi$ from repulsive to attractive DDI, the vortex-pattern formations obtained with 
$\varphi=0$ still survive up to some angle at which the DDI is attractive. 
Apart from the interesting stable vortex-pattern formations obtained with dipolar-symmetric and dipolar-asymmetric 
binary mixtures, one of the main outcome of the present study is verified by the almost-complete spatial separation in the 
two-component densities, which occurs for large polarization angles, when the DDI is attractive. 
We have verified half-space angular separations of the densities in the case of a dipolar-symmetric mixture, 
represented by $^{164}$Dy-$^{162}$Dy, whereas the separations have radial space distribution for the dipolar-asymmetric 
cases, represented by $^{168}$Er-$^{164}$Dy and $^{164}$Dy-$^{87}$Rb. In our study of the miscibility behavior with 
the polarization angle, we determine the critical polarization angles where the symmetry is broken, leading to the occurrence 
of an almost-complete space separation of the binary mixtures. 
As shown, the number of vortices and miscibility of the binary mixtures are significantly affected by the polarization, 
with the attractive part, obtained by increasing the angle $\varphi$, being relevant to reduce the radius of the coupled 
condensate, merging vortices together and establishing the observed almost-complete space separation.

Another quite relevant result of the present study is the observed effect of the mass-asymmetry in the 
vortex-pattern structures, which is evidenced in dipolar-symmetric binary mixtures.
As shown, the quite small mass-symmetry breaking given by the binary dipolar-symmetric mixture 
$^{164}$Dy-$^{162}$Dy is enough to modify significantly the two-species density patterns. 
This is evidenced by the results with $\delta=1$. For identical masses, 
instead of the set of patterns shown in Fig.~\ref{fig02}(C), one should verify 
identical patterns for both species shown at each particular polarization angle.
In view of our present conclusion on the mass-symmetry sensibility, 
when considering a purely mean-field approach for dipolar coupled systems, which could be confirmed experimentally
(following Ref.~\cite{2018arxiv-Ferlaino}) 
with binary mixtures having isotopes of the same kind of dipolar atoms,
we suggest to be of high interest further investigations on the role of quantum fluctuations and beyond mean-field properties 
of coupled dipolar condensates, following Refs.~\cite{2016-droplet,2016-Ferlaino}. 
The physics understanding behind self-bound droplet formation in single and binary dipolar mixtures can be improved 
by considering effects observed when tuning the dipole-dipole interactions. 
With dipolar-symmetric mixtures, droplet-like formations are more likely to occur due to the miscibility and 
competition between repulsive contact and attractive dipolar interactions when tuning the polarization angle.

\section*{Acknowledgements} 
RKK acknowledges the financial support from FAPESP of Brazil (Contract number 2014/01668-8). AG and LT thank 
CAPES, CNPq and FAPESP of Brazil for partial support. LT is also partially supported by
INCT-FNA (Proc. No. 464898/2014-5).

\end{document}